\documentclass[11pt,a4paper]{article}
\pdfoutput=1
\usepackage{jcappub}
\usepackage{rotating}
\usepackage{array}
\usepackage{amsmath}
\usepackage[normalem]{ulem}
\usepackage{slashed}
\usepackage{booktabs}
\usepackage[pdftex,table]{xcolor}
\usepackage{units}

\usepackage{multirow}

\arxivnumber{TTK-18-38}

\title{Freeze-in production of decaying dark matter in five steps}

\author{Saniya Heeba,}
\author{Felix Kahlhoefer}
\author{and Patrick St\"{o}cker}

\affiliation{Institute for Theoretical Particle Physics and Cosmology (TTK), RWTH Aachen University, \\ D-52056 Aachen, Germany}

\emailAdd{heeba@physik.rwth-aachen.de}
\emailAdd{kahlhoefer@physik.rwth-aachen.de}
\emailAdd{stoecker@physik.rwth-aachen.de}

\abstract{We study the cosmological evolution and phenomenological properties of scalar bosons in the keV to MeV range that have a tiny mixing with the Standard Model Higgs boson. The mixing determines both the abundance of light scalars produced via the freeze-in mechanism and their lifetime. Intriguingly, the parameters required for such scalars to account for all of the dark matter in the present Universe generically predict lifetimes comparable to the sensitivity of present and future indirect detection experiments. In order to accurately determine the relic abundance of light scalars, we calculate freeze-in yields including effects from finite temperatures and quantum statistics and develop a new approach for solving the Boltzmann equation for number-changing processes in the dark sector. We find that light scalars can potentially explain the anomalous x-ray emission at 3.5 keV, while evading constraints from structure formation and predicting potentially observable self-interaction cross sections.}


\begin{document}
\maketitle
\flushbottom

\section{Introduction}

All known particles of the Standard Model (SM) have sufficiently strong interactions to enter into thermal equilibrium with each other at high temperatures. It is hence tempting to assume that also the dark matter (DM) particle was at some point in thermal equilibrium with the SM and then obtained its relic abundance via the freeze-out mechanism. While this scenario is both well-motivated and predictive, there are many interesting alternative production mechanisms for DM. In particular, it is conceivable that reheating created only a negligible DM abundance and that the dark sector is only slowly populated subsequently through tiny interactions with SM particles. This so-called freeze-in mechanism~\cite{Hall:2009bx,Bernal:2017kxu} has received significant interest recently, illustrated best by the fact that \texttt{micrOmegas} now provides a numerical tool for the automated calculation of freeze-in yields~\cite{Belanger:2018ccd}.

While freeze-in is an attractive alternative to the standard paradigm of thermal freeze-out, an experimental confirmation of the idea is rather challenging. Colliders such as the LHC are typically unable to probe non-thermalised hidden sectors~\cite{Co:2015pka,Kahlhoefer:2018xxo} and direct detection experiments are only sensitive if the mediator of the interactions between the DM particle and the SM is light, such that scattering is strongly enhanced in the non-relativistic limit~\cite{Essig:2011nj,Chu:2011be,Hambye:2018dpi}. Indirect detection, however, may have a unique opportunity to probe freeze-in~\cite{Heikinheimo:2018duk}, for example if the DM particle is unstable, such that it can decay via the same interaction responsible for its production~\cite{Babu:2014pxa}. In this case, the smallness of the coupling implied by the freeze-in mechanism explains at the same time the long lifetime of DM in the absence of a stabilising symmetry.

Such a set-up is well-known in the context of keV sterile neutrinos, for which production and decay both proceed through a tiny mixing with SM neutrinos~\cite{Dodelson:1993je} (see also Refs.~\cite{Kang:2014mea,Shakya:2015xnx,Konig:2016dzg} for alternative ways to produce sterile neutrinos via freeze-in). The case of scalar and vector particles that mix with the SM Higgs boson or the SM photon, respectively, was first explored in Ref.~\cite{Berger:2016vxi}.\footnote{The case of \emph{stable} scalar singlets produced via the Higgs portal, was considered previously in Refs.~\cite{Yaguna:2011qn,Blennow:2013jba,Heikinheimo:2016yds}.} This work however focused on the case of a sub-dominant DM component with a lifetime short compared to the age of the Universe. Moreover, the freeze-in production of light scalars was calculated in a very approximate way, considering only a limited range of temperatures and a few production processes.

In the present work we point out that the production in the Early Universe of light scalars with Higgs mixing is in fact surprisingly complex. First of all, the freeze-in production proceeds in three stages, corresponding to temperatures before, during and after electroweak symmetry breaking (EWSB). Although the dominant contribution to the DM abundance typically arises after EWSB, it is essential to correctly account for the effect of the electroweak phase transition (EWPT) in order to avoid unphysical contributions from high temperatures. Furthermore, since the light scalars are not protected by a stabilising symmetry, $2\to3$ and $3\to2$ processes may play an important role. This means that the co-moving DM density is not necessarily constant after the end of freeze-in and additional considerations are needed to calculate the subsequent evolution of the dark sector.

A number of previous works have studied the freeze-out of number-changing processes in a dark sector that is initially in kinetic and chemical equilibrium~\cite{Carlson:1992fn,Bernal:2015bla,Pappadopulo:2016pkp,Farina:2016llk}. In the present work, we extend these studies by considering the evolution of a dark sector where chemical equilibrium is not guaranteed.\footnote{Refs.~\cite{Heikinheimo:2016yds,Heikinheimo:2017ofk} provide a similar discussion in the context of $2\to4$ processes, while Ref.~\cite{Bernal:2015ova} considers $2\to3$ processes for the case of vector DM.} For this purpose, we consider $2\to3$ processes for relativistic initial states in order to address the question whether or not number-changing processes are efficient enough to thermalise the dark sector. Using a combination of analytical approximations and numerical algorithms we can then calculate the present day abundance of light scalars for arbitrary model parameters.

\begin{figure}
\centering
	\includegraphics[width=.8\textwidth]{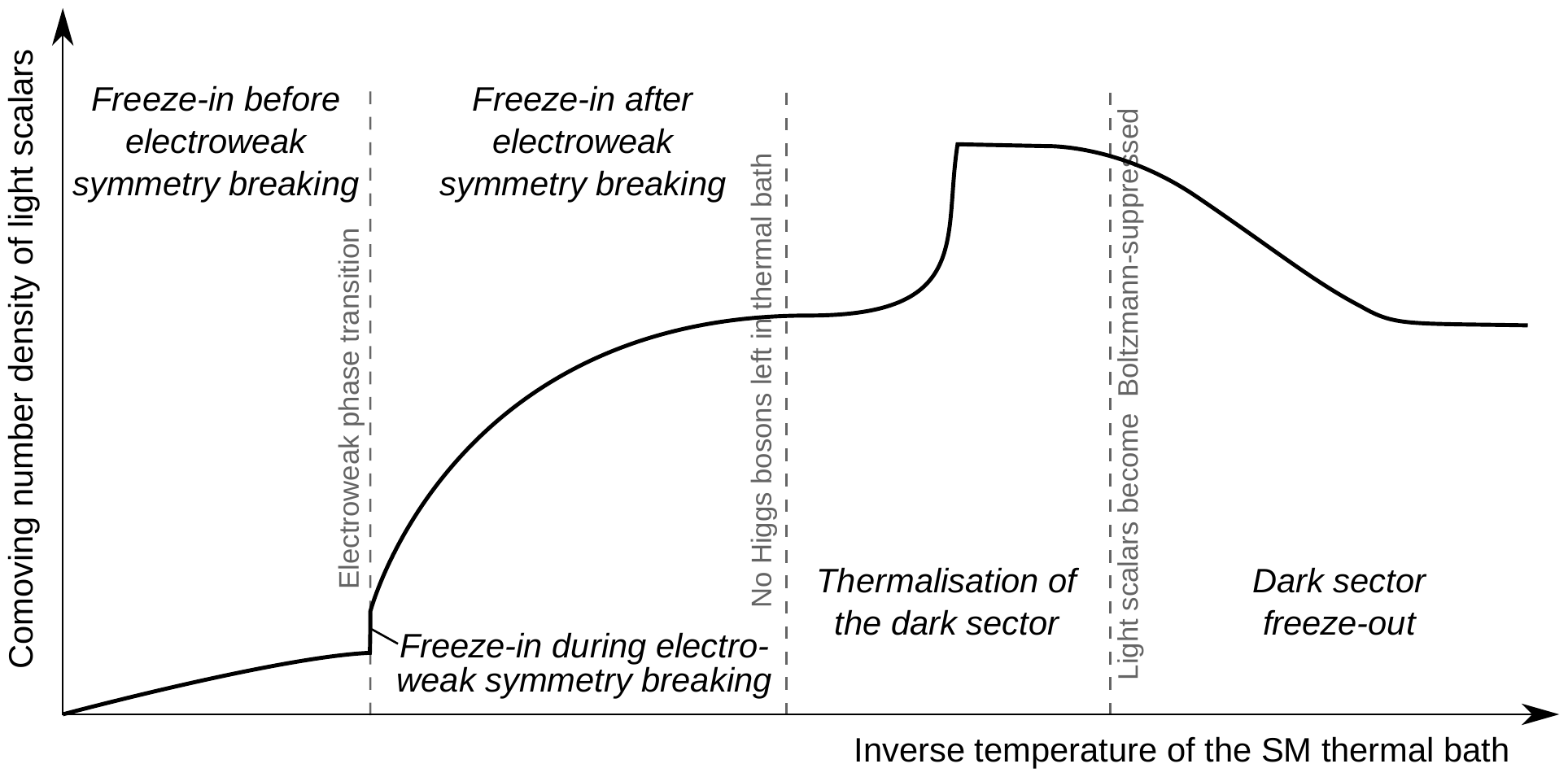}
\caption{Sketch of the evolution of the co-moving number density of light scalars with decreasing temperature of the SM thermal bath. Note that the relative size of the different regimes and the magnitude of the different effects is not to scale.\label{fig:sketch}}
\end{figure}

The different stages of production of light scalars are illustrated in figure~\ref{fig:sketch}, which shows schematically the evolution of the co-moving number density $Y$ as a function of the inverse temperature $x$ (precise definitions of these quantities will be provided below). The first three steps correspond to the freeze-in production of light scalars before, during and after EWSB. The fourth and fifth step correspond to thermalisation of the dark sector and dark sector freeze-out, respectively. While the first three steps always increase the DM abundance and the final step always leads to a decrease, the fourth step can either enhance or deplete the co-moving number density, depending on how the freeze-in yield compares to the equilibrium distribution.

We identify large regions of parameter space where lights scalars can be all of DM and have a lifetime large compared to the age of the Universe. The combined constraints from indirect detection experiments and DM self-interactions, however, force such light scalars to have masses in the keV range and lifetimes that are only slightly below current experimental sensitivity, such that the scenario may be testable with future x-ray missions such as the Hitomi~\cite{Aharonian:2016gzq} re-flight. Moreover, decaying light scalars from freeze-in may provide a viable explanation of the claimed observation of an x-ray line at 3.5 keV in various astrophysical systems~\cite{Bulbul:2014sua,Boyarsky:2014jta} while facing much weaker constraints from structure formation than sterile neutrinos. For different ways of connecting the 3.5 keV line to the freeze-in mechanism, we refer to Refs.~\cite{Brdar:2017wgy,Heeck:2017xbu}.

This paper is structured as follows. We present the model that we consider in section~\ref{sec:model}, derive the mixing with the SM Higgs boson after electroweak symmetry breaking and discuss the importance of finite-temperature effects. Section~\ref{sec:freeze-in} deals with the freeze-in production of light scalars, accounting for the different production mechanisms before, during and after electroweak symmetry breaking. The subsequent evolution, in particular the role of $2\to3$ processes, is discussed in section~\ref{sec:evolution}, where we also provides the final results from the relic density calculation. The phenomenological implications and existing experimental constraints are discussed in Sec.~\ref{sec:pheno}. Appendix~\ref{app:boltzmann} provides additional details on how to solve the Boltzmann equation for $2\to3$ processes.

\section{Light scalars and Higgs mixing}
\label{sec:model}

We consider the most general Lagrangian for a real scalar singlet $s_0$ with a $\mathbb{Z}_2$ symmetry (see also Ref.~\cite{Evans:2017kti}):
\begin{align}
 \mathcal{L}_\text{scalar} & = \frac{1}{2} \partial^\mu s_0 \partial_\mu s_0 + \frac{1}{2} \mu_s^2 \, s_0^2 - \frac{1}{4} \lambda_s \, s_0^4 - \frac{1}{2} \lambda_{hs} \, |H|^2 s_0^2  \; .
\end{align}
The negative mass term for the scalar leads to a spontaneous breaking of the $\mathbb{Z}_2$, such that the scalar acquires a vacuum expectation value (vev): $s_0 = s + v_s$. We assume for the moment that this happens at temperatures large compared to the temperature of the EWPT. The Lagrangian then becomes
\begin{align}
 \mathcal{L}_\text{scalar} & = \frac{1}{2} \partial^\mu s \partial_\mu s - \frac{1}{2} m_s^2 \, s^2 - \frac{1}{4} \lambda_s \, s^4 - \lambda_s \, v_s \, s^3 - \frac{1}{2} \lambda_{hs} \, |H|^2 (s^2 + 2 \, s \, v_s) \; .
\end{align}
Here we have assumed that $\lambda_{hs} \, v_s^2$ is very small compared to the bare Higgs mass term and can therefore be neglected, which is a good approximation for the values of $\lambda_{hs}$ and $v_s$ that we will consider later.

After EWSB we replace $H = (h + v)/\sqrt{2}$. This introduces an additional contribution to the mass term of the scalar singlet: $-\mu_s^2 \to -\mu_s^2 + \lambda_{hs} v^2 / 2$. This contribution can potentially restore the $\mathbb{Z}_2$ symmetry, leading to the so-called vev flip-flop~\cite{Baker:2016xzo,Baker:2017zwx}. Here, instead, we will consider the case where the $\mathbb{Z}_2$ remains broken at low temperatures. In this case, the Lagrangian can be written as
\begin{align}
 \mathcal{L}_\text{scalar} & = \frac{1}{2} \partial^\mu s \partial_\mu s - \frac{1}{2} m_s^2 \, s^2 - \frac{1}{4} \lambda_s \, s^4 - \lambda_s \, v_s \, s^3 - \frac{1}{4} \lambda_{hs} \, (h^2 + 2 \, h \, v) (s^2 + 2 \, s \, v_s) \label{eq:lagrangian} \; .
\end{align}
We take the free parameters to be $\lambda_s$ and $\lambda_{hs}$ as well as the zero-temperature value of $m_s$, such that the vev of the scalar field is given by
\begin{align}
 v_s & = \frac{m_s}{\sqrt{2 \, \lambda_s}} \; .
\end{align}
Eq.~(\ref{eq:lagrangian}) still contains a non-diagonal mass term of the form $\mathcal{L}_\text{mixing} = \lambda_{hs} \, v_s v \, \, h s$, leading to the mixing of the two scalar fields. The mass matrix is given by
\begin{equation}
M = \begin{pmatrix}
2 v^{2 }  \lambda & \lambda_{hs} v_{s} v \\ \lambda_{hs} v_{s} v & 2 v_{s}^{2} \lambda_{s} 
\end{pmatrix}
\end{equation}
with $\lambda$ denoting the quartic self-coupling of the SM Higgs field.
To find the mass eigenstates we rotate the interaction eigenstates by the mixing angle $\theta$:
\begin{equation}
 \begin{pmatrix}h_\text{SM} \\ h_s\end{pmatrix} = \begin{pmatrix}\cos \theta & \sin \theta \\ -\sin \theta & \cos \theta \end{pmatrix} \begin{pmatrix}h \\ s\end{pmatrix} \;.
\end{equation}
The mass matrix is diagonalised for
\begin{equation}
 \tan  2 \theta \equiv \frac{\lambda_{hs} \, v \, v_s}{\lambda \, v^2 - \lambda_s \, v_s^2 } \; ,
 \label{eq:tan2theta}
\end{equation}
which gives the mass eigenvalues
\begin{align}
m^{2}_{h,s} = \lambda v^{2} + \lambda_{s} v_{s}^{2} \pm \sqrt{(\lambda v^{2} - \lambda_{s} v_{s}^{2})^{2} + \lambda_{hs}^{2} v_{s}^{2} v^{2}}\;.
\end{align}
We will focus on the case $\theta \ll 1$, such that $h_\text{SM}$ is SM-like, while $h_s$ only has tiny couplings to SM states. In this case we can furthermore neglect the shift in the mass eigenstates.\footnote{In particular, the relative change of $m_s$ will be of order $\lambda_{hs}^2 / \lambda_{s}$, which is completely negligible.}
Thus, to simplify notation, we will continue denoting the mostly SM-like scalar by $h$ and the mostly SM-singlet scalar by $s$.
We also note that in this case eq.~(\ref{eq:tan2theta}) simplifies to
\begin{equation}
\theta = \frac{\lambda_{hs} \, v \, v_s}{m^{2}_{h} - m^{2}_{s}}\;. 
\end{equation}

While the discussion above has been fully general, we will from now on focus on the case where the extra scalar is much lighter than the SM Higgs boson, $m_s \ll m_h$. The main motivation for this choice is that we will be interested in the case of long-lived light scalars, which~-- as we will see below~-- requires masses in the keV to MeV range.

We make the crucial observation that the mixing angle $\theta$ depends on temperature via the temperature dependence of the two vevs. In particular, $\theta = 0$ for $v = 0$, i.e.\ before EWSB. This means that the phenomenology of our model will be decisively different at high temperatures, when the two scalar fields cannot mix, and at low temperatures, when mixing becomes possible. We will treat these two regimes separately below and also discuss in detail what happens for temperatures close to the EWPT.

The temperature dependence of the electroweak vev can be determined by considering the effective potential $V$. Assuming the additional scalar field to be so weakly coupled that its contribution is completely negligible, we recover the well-known result for the SM~\cite{Quiros:1999jp}:
\begin{equation}
 V(\phi,T) = D(T^2 - T_o^2) \phi^2 - E T \phi^3 +\frac{\lambda(T)}{4}\phi^4 \; ,
\end{equation}
where $\phi$ is the constant background field and $D, E, T_o$ and $\lambda(T)$ can be calculated in terms of SM parameters. Since we are not interested in the strength of the EWPT, we will ignore the term proportional to $\phi^3$, making use of the fact that $E$ is known to be small in the SM. The temperature of the EWPT is then simply given by $T_\mathrm{c} = T_o \approx 164\,\mathrm{GeV}$. For $T > T_\mathrm{c}$ the effective potential is minimized by $\phi = 0$, such that the electroweak symmetry is unbroken. While fermions and gauge bosons are massless, the mass of the (complex) Higgs boson is given by
\begin{equation}
 m_H(T)^2 = D (T^2 - T_o^2) \; .
\end{equation}
For temperatures smaller than but close to the critical temperature, the Higgs vev and mass are simply given by
\begin{align}
 v(T)^2 & = \frac{2 D (T_o^2 - T^2)}{\lambda(T)} \; ,\\
 m_h(T)^2 & = 2 \lambda(T) v(T)^2 = 4 D (T_o^2 - T^2) \; .
\end{align}
We will make use of these expressions in the following section.

\section{Freeze-in production of light scalars}
\label{sec:freeze-in}

We will be interested in the case where the initial abundance of light scalars (e.g.\ at the end of reheating) is completely negligible, such that the dominant contribution to their relic abundance stems from the ``leakage'' of energy from the visible into the dark sector. There are three relevant epochs to this freeze-in production, corresponding to the temperature-dependent properties of the SM Higgs boson. As long as the electroweak symmetry is unbroken, the SM Higgs boson and the light scalar cannot mix and hence only a small number of production channels are available. During\footnote{Strictly speaking, oscillations happen immendiately \emph{after} the EWPT, but the relevant temperatures are very close to $T_\mathrm{c}$.} the EWPT, the mixing between the two scalar bosons can be large for a very short period of time, potentially leading to rapid production of light scalars via oscillations. At lower temperatures the mixing angle becomes small but non-zero, opening up a wide range of possible production modes.

\subsection{Step 1: Production before EWSB}
\label{sec:beforeEWSB}

\begin{figure}
\centering
	\includegraphics[scale=0.7,clip,trim = 20 50 00 70]{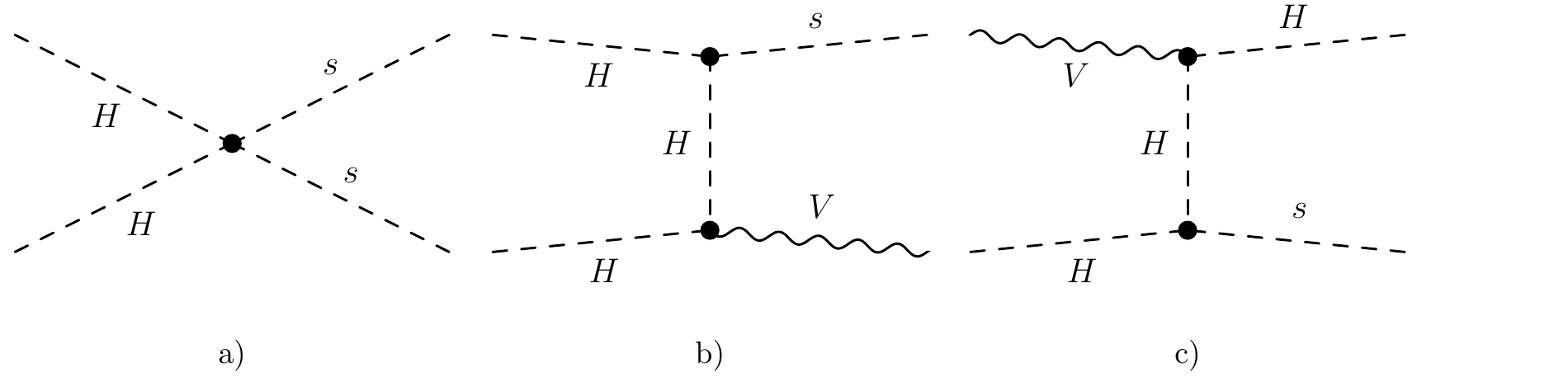}
\caption{\label{fig:beforeEWSB} Processes relevant for the freeze-in production of light scalars before electroweak symmetry breaking, with $V$ denoting a gauge boson of $SU(2)$ or $U(1)$. The four-point interaction (a) gives a larger contribution than diagrams with a $t$-channel Higgs boson (b), (c).}
\end{figure}

Before EWSB, the SM Higgs field has vanishing vev and therefore cannot mix with the light scalar. This means that there are only two types of processes that can contribute to the freeze-in production of light scalars: the four-point interaction $H^\dagger H s^2$ and diagrams with a complex Higgs boson in the $t$-channel, for example $H^\dagger H \to s V$ or $HV \to H s$, with $V$ being a (massless) gauge boson of either $SU(2)$ or $U(1)$ (see figure~\ref{fig:beforeEWSB}). Conversely, diagrams like $t \bar{t} \to g s$ with a fermion in the $t$-channel are absent, because the light scalar does not at this stage couple to top quarks. We therefore find that all relevant cross sections vanish in the limit $\sqrt{S} \to \infty$, such that freeze-in production proceeds dominantly at low temperatures. This means in particular that in the model we consider the freeze-in yield is independent of the reheating temperature $T_\mathrm{r}$, as long as $T_\mathrm{r} \gg T_\mathrm{c}$. 

Since the mixing angle $\theta$ vanishes before EWSB and since the mass of the light scalar can be neglected at high temperatures, the only parameters that determine the yield from freeze-in before EWSB are the mixed quartic coupling $\lambda_{hs}$ and the vev $v_s$ of the light scalar. We find that processes involving a $t$-channel Higgs boson are completely negligible, so that the freeze-in yield is dominated by the process $H^\dagger H \to s s$. To leading order in $\lambda_{hs}$ the corresponding cross section as a function of the centre-of-mass energy $\sqrt{S}$ is given by 
\begin{align}
 \sigma_{H^\dagger H \to ss} & = \frac{\lambda_{hs}^{2}}{32 \pi  S} \frac{(S+2m_s^2)^2}{(S-m_s^2)^2}\sqrt{\frac{ S - 4m_s^2}{ S-4m_H^2}} \approx \frac{\lambda_{hs}^{2}}{32 \pi \sqrt{ S^2-4\,S\,m_H^2}} \; ,
\end{align}
which is in fact independent of $v_s$.

The number density of light scalars $n_s$ can then be obtained by solving the Boltzmann equation
\begin{align}
\label{eq:BoltzmannGeneral}
\dot{n}_{s} + 3 \, H \, n_{s} = C_H \, g_{H} \int C[f_{H}] \frac{\mathrm{d}^{3}p}{(2 \pi)^{3}} \; .
\end{align}
Here $g_H = 2$ denotes the degrees of freedom of the Higgs field, $C_H = 2$ reflect the fact that there are two non-identical particles in the initial state and $H$ is the Hubble expansion rate during radiation domination:
\begin{align}
H = \sqrt{g_\mathrm{SM}^\ast} \frac{1.66 \, T_\mathrm{SM}^{2}}{M_\mathrm{Pl}}
\end{align}
with $M_\mathrm{Pl}$ being the Planck mass and $T_\mathrm{SM}$ and $g^\ast_\mathrm{SM}$ denoting the temperature and number of relativistic degrees of freedom of the thermal bath of SM particles. 

The quantity $C[f_H]$ represents the collision term responsible for creating light scalars from the thermal bath. To obtain a simple expression, we introduce the dimensionless quantities, $Y = n/s_\mathrm{SM}$  and $x_\mathrm{SM} = m_s/T_\mathrm{SM}$, where $s_\mathrm{SM}$ is the entropy density of the visible sector:
\begin{align}
 s_\mathrm{SM}  = g_\mathrm{SM}^\ast \frac{2 \pi^{2} \, T_\mathrm{SM}^{3}}{45} \; .
\end{align}
We can then use the methods outlined in Ref.~\cite{Hall:2009bx} to simplify eq.~(\ref{eq:BoltzmannGeneral}) and obtain
\begin{align}
\frac{\mathrm{d}Y_s}{\mathrm{d}x_\mathrm{SM}} & =  \frac{2 \, s_\mathrm{SM}}{H  \, x_\mathrm{SM}} \langle\sigma v\rangle \, Y_H^2 \; .
\end{align}

\begin{figure}
\centering
	\includegraphics[scale=0.7]{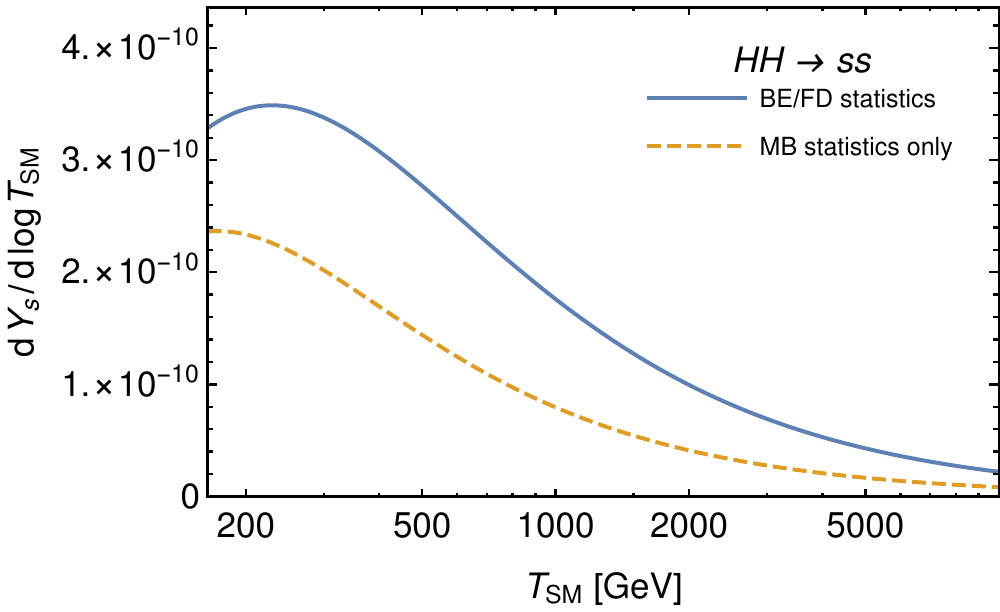}
\caption{\label{fig:stats_beforeEW}Differential yield of light scalars before EWSB as a function of temperature for the dominant production channel $H^\dagger H\to ss$. The blue solid line corresponds to the result obtained when accounting for quantum statistics (i.e.\ using Bose-Einstein and Fermi-Dirac distributions), while the orange dashed line represents the results obtained when assuming a Maxwell-Boltzmann distribution. For this plot we have set $\lambda_{hs} = 2 \times 10^{-10}$.}
\end{figure}

Calculating the thermally averaged cross-section $\langle\sigma v\rangle$ in a similar fashion as in Ref.~\cite{GONDOLO1991145}, we find the differential production rate of DM to be given by 
\begin{align}
\frac{\mathrm{d}Y_s}{\mathrm{d}x_\mathrm{SM}} & = \frac{90 \, M_\mathrm{Pl}}{(2 \pi)^6 \, 1.66 \, (g_\mathrm{SM}^\ast)^{3/2}} \frac{x^3}{m_s^{4}} \int_{S_{min}}^{\infty} \mathrm{d}S \, (S - 4 \, m_H^{2}) \, \sigma \, \sqrt{S} \, K_{1}\left(\sqrt{S } \ x/m_s\right) \; ,
\label{eq:beforeEWSB}
\end{align}
where $K_1(x)$ is the Modified Bessel function. We show $\mathrm{d} Y_s / \mathrm{d} \log T_\mathrm{SM}$ as a function of $T_\mathrm{SM}$ in figure~\ref{fig:stats_beforeEW}. As expected for freeze-in via renormalisable interactions, we find that the production rate peaks at low temperatures, i.e.\ right before the EWPT. Eq.~(\ref{eq:beforeEWSB}) can be immediately integrated from $x = 0$ to $x = m_s / T_\mathrm{c}$ to obtain the total amount of DM production before EWSB.

We find that for $m_s \ll T_\mathrm{c}$ the number of DM particles produced before EWSB is to good approximation independent of the scalar mass and can be written as
\begin{align}
\label{eq: Y1}
\Delta Y_{s,1} \approx (3.58 \times 10^{9}) \  \lambda_{hs}^2 \; .
\end{align}
We emphasize that this result consistently takes into account the fact that the mass of the Higgs bosons in the initial state depends on temperature. However, since $m_H(T) < T$ before EWSB, it is not a good approximation to treat the Higgs bosons in the initial state as non-relativistic, which we have done implicitly by using Maxwell-Boltzmann statistics to calculate the thermally-averaged annihilation cross section. 

In order to properly account for the statistics of indistinguishable relativistic particles, we use \texttt{micrOmegas}~\cite{Belanger:2018ccd}, which employs Bose-Einstein statistics to calculate the relic abundance more accurately. It is however not easily possible to implement a temperature-dependent Higgs mass in \texttt{micrOmegas}. We therefore determine the value of the Higgs boson mass that gives the same result as the temperature-dependent Higgs mass for the case of Maxwell-Boltzmann statistics, and then fix the Higgs boson mass to this value in \texttt{micrOmegas} also for the case of Bose-Einstein statistics. The suitable value is found to be $m_{H,\text{fixed}} \approx m_H(T = 287\,\mathrm{GeV}) \approx 136 \,\mathrm{GeV}$, which is consistent with the observation that the dominant contribution to freeze-in production arises shortly before the EWPT.

Following this procedure, we find that the abundance with the more accurate treatment of statistics is given by
\begin{align}
\Delta Y_{s,1} ^{\text{BE}} \approx (6.77\times 10^{9}) \ \lambda_{hs}^2 \; ,
\end{align}
which is larger than the previous result by about a factor of 2 (see figure~\ref{fig:stats_beforeEW}). Such a difference is not unexpected for relativistic particles in the initial state and is consistent with similar observations made in Ref.~\cite{Belanger:2018ccd}.

\subsection{Step 2: Production during EWSB}

Electroweak symmetry breaking is triggered by the $\mu^2$-term in the Lagrangian changing sign. At the point when this happens, $T \approx T_\mathrm{c}$, the mass of the SM Higgs boson vanishes and then increases again with decreasing temperature. This implies in particular that right after the EWPT the SM Higgs boson will have both a very small (but non-zero) vev $v(T)$ and a very small mass $m_h(T)$. The temperature-dependent mixing angle is then approximately given by
\begin{equation}
 \theta(T) = \frac{\lambda_{hs} \, v(T) \, v_s}{m_h^2(T) - m_s^2} \; ,
 \label{eq:thetaofT}
\end{equation}
where we assume for simplicity that the temperature dependence of $m_s$ and $v_s$ is negligible for $T \approx T_\mathrm{c}$.

For a short period of time right after the EWPT, the mass of the SM Higgs boson may be comparable to the mass of the light scalar, $m_h(T) \approx m_s$. It then becomes important to include also the imaginary part of the Higgs boson mass, which depends on the interactions of the Higgs with other particles in the thermal bath. While this term ensures that $\theta(T) \ll 1$ (assuming sufficiently small $\lambda_{hs}$), the mixing between the two scalars will be substantially enhanced. At this point, SM Higgs bosons can directly be converted into light scalars via oscillations.

This production mechanism is analogous to the one of hidden photons via mixing with the SM photon, which is enhanced at finite temperatures and densities due to plasma effects. It was shown in Ref.~\cite{Redondo:2008ec} that the resulting abundance of hidden photons depends essentially only on the mixing parameter and on how quickly the mass of the visible photon varies with temperature. The corresponding result for our case can be written as
\begin{equation}
 \Delta Y_{s,2} = \frac{\pi \, \zeta(2) \, Y_h(T_\mathrm{osc})}{\zeta(3)}  \frac{\lambda_{hs}^2 \, v(T_\mathrm{osc})^2  \, v_s^2}{H(T_\mathrm{osc}) \, T_\mathrm{osc}^2} \left| \frac{\mathrm{d} m_h^2}{\mathrm{d}T}\right|^{-1}_{T = T_\mathrm{osc}} \; ,
\end{equation}
where $T_\mathrm{osc}$ is the temperature for which $m_h(T) = m_s$. Since $T_\mathrm{osc} \approx T_\mathrm{c}$, we define $\epsilon = (T_\mathrm{c} - T_\mathrm{osc}) / T_\mathrm{c} \ll 1$.
We then find $m_h^2 \approx 8 \,D \, T_\mathrm{c}^2 \, \epsilon$ and hence 
\begin{equation}
 \epsilon = \frac{m_s^2}{8\,D\,T_\mathrm{c}^2} \; .
\end{equation}
For $\epsilon \ll 1$ we find furthermore
\begin{equation}
 v(\epsilon)^2 = 4 \ T_\mathrm{c}^2 \frac{D \, \epsilon}{\lambda(T_\mathrm{c})}
\end{equation}
and hence
\begin{equation}
 \frac{\lambda_{hs}^2 \, v(T_\mathrm{osc})^2  \, v_s^2}{H(T_\mathrm{osc}) \, T_\mathrm{osc}^2} \left| \frac{\mathrm{d} m_h^2}{\mathrm{d}T}\right|^{-1}_{T = T_\mathrm{osc}} \approx \frac{\lambda_{hs}^2 \, m_s^2 \, v_s^2}{16 \, D \,  \lambda(T_\mathrm{c}) \, H(T_\mathrm{c}) \, T_\mathrm{c}^3} \; .
\end{equation}
We can furthermore make use of the fact that the Higgs boson is highly relativistic at $T = T_\mathrm{osc}$ and hence
\begin{equation}
 \frac{\pi \, \zeta(2) \, Y_h(T_\mathrm{osc})}{\zeta(3)} \approx \frac{\zeta(2) \, T_\mathrm{c}^3}{\pi \, s_\text{SM}(T_\mathrm{c})} \; .
\end{equation}
Bringing everything together, we obtain for the DM yield from oscillations
\begin{equation}
\label{eq: Y2}
 \Delta Y_{s,2} \approx (1.93 \times 10^5 \, \mathrm{GeV^{-4}}) \, \lambda_{hs}^2 \, m_s^2 \, v_s^2 \; .
\end{equation}
For the parameter region that we will be interested in ($m_s \ll v_s \sim \mathrm{GeV}$) this yield is small compared to the one from before EWSB, such that the final DM abundance does not change significantly during EWSB.

\subsection{Step 3: Production after EWSB}
\label{sec:afterEWSB}

\begin{figure}
\centering
	\includegraphics[scale=0.9,clip,trim = 20 60 180 70]{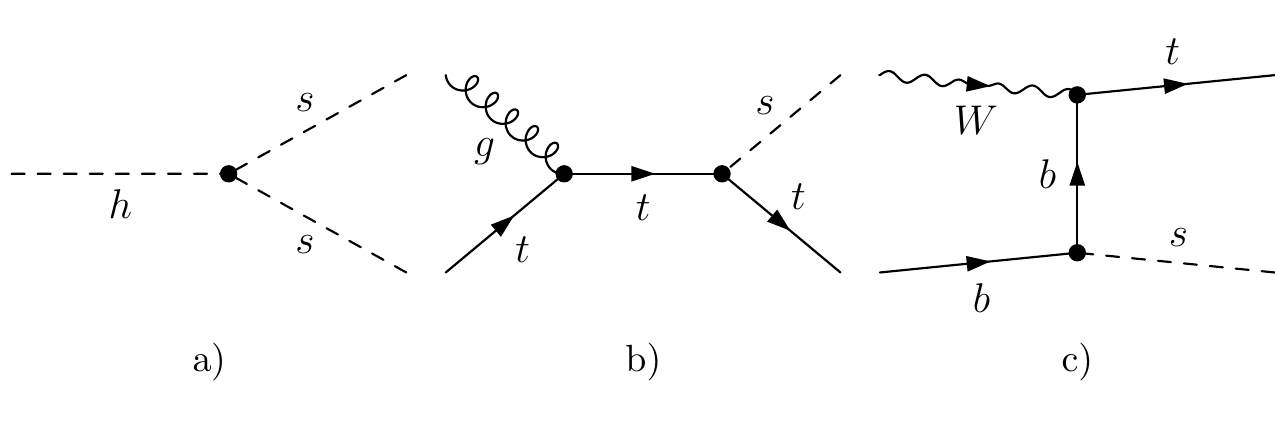}
\caption{\label{fig:afterEWSB}Processes relevant for freeze-in production after the electroweak phase transition. For $v_s < 100 \,\mathrm{GeV}$, production is dominated by Higgs decays (a). Conversely, for $v_s > 100 \,\mathrm{GeV}$, a number of diagrams like the ones shown in b) and c) give relevant contributions.}
\end{figure}

As the temperature decreases further, oscillations between SM Higgs bosons and the light scalars become suppressed, but a number of new production channels open up (see figure~\ref{fig:afterEWSB} for a few examples). First of all, light scalars can now be produced in decays of SM Higgs bosons: $h \to s s$. Moreover, processes like $t \bar{t} \to s g$ now become active and contribute to the abundance of light scalars. We note that, in contrast to the $t$-channel processes considered in Sec.~\ref{sec:beforeEWSB}, these processes grow with energy like $\log ( S / m_t^2 )$, due to the divergent contribution in the collinear limit $\cos \theta \to 0$. In the absence of a phase transition at high temperatures, these processes would therefore dominate the production of light scalars. However, since these processes become active only after EWSB, their contribution is less important.

\begin{figure}
\centering
\includegraphics[scale=0.7]{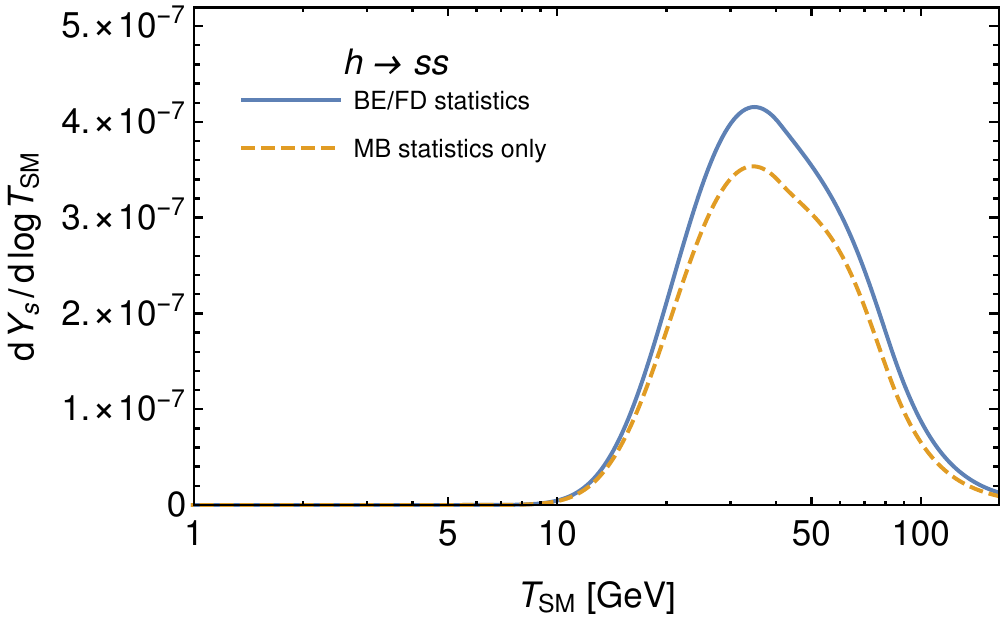}\qquad
\includegraphics[scale=0.7]{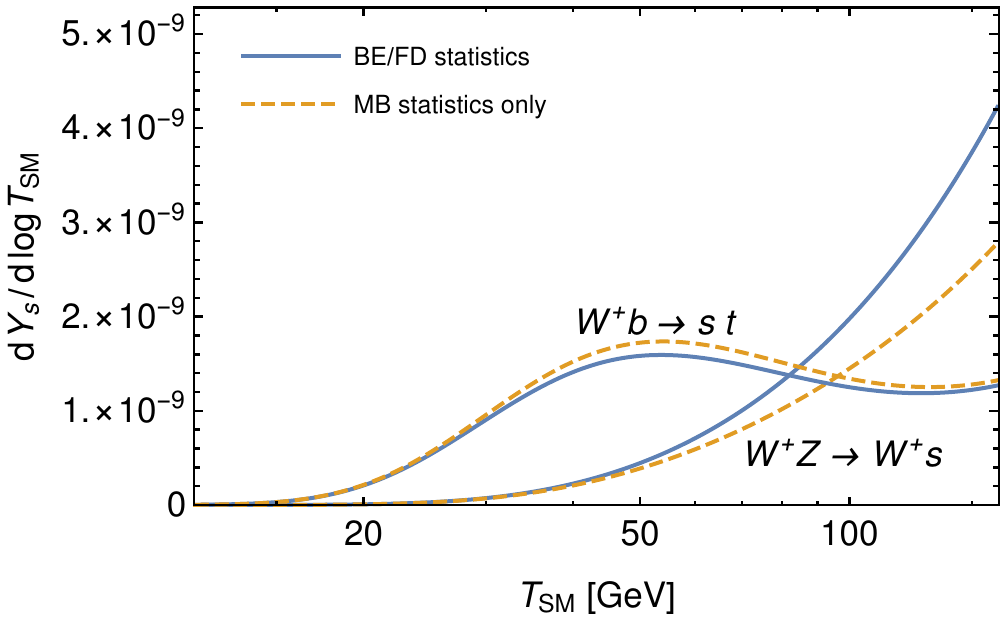}
 \caption{\label{fig:stats_afterEW} Differential yield of light scalars after EWSB as a function of temperature for a number of different production channels. Blue solid lines correspond to the results obtained when accounting for quantum statistics (i.e.\ using Bose-Einstein and Fermi-Dirac distributions), while orange dashed lines represent the results obtained when assuming a Maxwell-Boltzmann distribution. For this plot we have set  $\lambda_{hs} = 2 \times 10^{-10}$ and $v_s = 25\,\mathrm{GeV}$. For these parameters, the dominant contribution arises from the decay $h\to ss$, which peaks around $T_\text{SM} \sim 40\,\mathrm{GeV}$.}
\end{figure}

Interestingly, the two types of processes depend in different ways on the fundamental parameters. The production from Higgs boson decays is proportional to $\lambda_{hs} \, v \propto \sin \theta / v_s$, whereas the production from SM fermions is simply proportional to $\sin \theta$. Hence, by independently varying $\lambda_{hs}$ and $v_s$, it is possible to divide the parameter space into two regions  on the basis of which channel(s) gives the dominant contribution.  We find that for $v_s \leq 100 \ \mathrm{GeV}$, or equivalently $\lambda_s \geq 5 \times 10^{-5} (m_s / 1 \, \mathrm{GeV})^2$, the dominant contribution ($ \geq 50 \%$) comes from Higgs decays:
\begin{align}
\Gamma_{h\rightarrow ss} = \frac{\sin^2 \theta}{v_s^2}\ \frac{\sqrt{m_h^2 -  4 m_s^2} \ (m_h^2+2m_s^2)^2}{32\pi \ m_h^2} \; .
\end{align}
The corresponding yield can be calculated using a simplified Boltzmann equation~\cite{Hall:2009bx}:
\begin{align}
\frac{\mathrm{d}Y^{\mathrm{decay}}_{s,3}}{\mathrm{d}x} = \frac{45 \, M_\mathrm{Pl}}{2 \pi^4 \, 1.66 \, (g_\mathrm{SM}^\ast)^{3/2}} \frac{m_h^2}{m_s^4} \, \Gamma_{h \rightarrow ss} \ x^3 K_1\left(m_h \, x / m_s\right)  \; .
\label{eq:Higgs_decay}
\end{align}
We show $\mathrm{d} Y_s / \mathrm{d} \log T_\mathrm{SM}$ as a function of $T_\mathrm{SM}$ in figure~\ref{fig:stats_afterEW}. Integrating this expression from $x = m_s / T_\mathrm{c}$ to $x \to \infty$, we obtain
\begin{align}
\label{eq: Y3a}
\Delta Y^{\mathrm{decay}}_{s,3} \sim \left(9 \times 10^{15} \ \mathrm{GeV}^{2}\right) \ \frac{\sin^2 \theta}{v_s^2} \ \sim 2.2 \times 10^{12} \ \lambda_{hs}^2\; .
\end{align}

Conversely, in the opposite regime ($v_s \geq 100 \ \mathrm{GeV}$), production is divided among a number of channels, with each channel contributing less than $10 \%$ to the total yield.  We perform the relevant calculations using  \texttt{micrOmegas} which identifies more than 100 possible channels. For reference, one such channel is $W^+b \rightarrow ts$ with a cross section proportional to $\sin^2 \theta$. In the limit $m_s \ll m_h$, the total yield from these processes is calculated to be\footnote{Our result differs from the one obtained in Ref.~\cite{Berger:2016vxi}, where only two production channels ($t\bar{t} \to s g$ and $tg\to ts$) are included. Moreover, Ref.~\cite{Berger:2016vxi} appears not to correctly take into account the temperature dependence of the production rate, which peaks at $T \gg m_s$ (see figure~\ref{fig:stats_afterEW}).}
\begin{align}
\label{eq: Y3b}
\Delta Y^{2\rightarrow2}_{s,3} \sim 7.2\times 10^{11} \, \sin^2 \theta \ \sim \left(1.8 \times 10^{8} \ \mathrm{GeV}^{-2}\right)\, \lambda_{hs}^2 \, v_s^2\; .
\end{align}

We also find that including quantum statistics affects the two regimes differently. In case of decay-dominated production, including quantum statistics increases the yield by approximately $20\%$. In the second regime, the resulting change is less than $1\%$. The reason is that a large number of different channels contribute, which behave in contrary ways on inclusion of quantum statistics. For example, the yield from channel $W^+Z\rightarrow W^+s$ increases whereas that from $W^+b \rightarrow ts$ decreases (see figure~\ref{fig:stats_afterEW}), so the net effect is washed out.

\bigskip

We can now calculate the total yield from freeze-in by combining the results obtained above:
\begin{equation}
 Y_{s,\text{fi}} = \Delta Y_{s,1} + \Delta Y_{s,2} + \Delta Y_{s,3}^\text{decay} + \Delta Y_{s,3}^{2\to2}
\end{equation}
From eqs.~(\ref{eq: Y1}), (\ref{eq: Y2}), (\ref{eq: Y3a}) and (\ref{eq: Y3b}), it is easy to see that DM production is dominated by processes after EWSB, i.e. $Y_{s,\text{fi}} \approx Y_{s,3}$. For $v_s  \gg 100 \ \mathrm{GeV}$, the contributions from step 1 and 2 could be comparable but remain smaller than that of step 3. 

Under the assumption that the initial DM yield after reheating is negligible compared to yield from freeze-in, we obtain $Y_s = Y_{s,\text{fi}}$, which can be directly converted into the present-day relic abundance $\Omega_s$:
\begin{equation}
 \Omega_s = \frac{m_s \, s_{\mathrm{SM},0}}{\rho_{\mathrm{c},0}} Y_{s} \; ,
\end{equation}
where $s_{\mathrm{SM},0}$ and $\rho_{\mathrm{c},0}$ denote the present-day entropy density and present-day critical density, respectively.\footnote{Here we assume implicitly that the scalars have a lifetime that is large compared to the age of the Universe. We will revisit this assumption in Sec.~\ref{sec:pheno}.} We can then compare the observed value $\Omega_\text{DM} h^2 = 0.12$~\cite{Aghanim:2018eyx} to the prediction in order to determine the value of $\lambda_{hs}$ required for light scalars to constitute all of DM. For the case of dominant production via Higgs decays we then obtain
\begin{equation}
 \lambda_{hs} = 4.0 \times 10^{-10} \left(\frac{m_s}{1\,\mathrm{MeV}}\right)^{-1/2} \; .
 \label{eq:lambdahs}
\end{equation}
These results can now be used as input for studying the further evolution of the dark sector.

Before doing so let us briefly revisit our assumption that the $\mathbb{Z}_2$ symmetry of the light scalar is already broken before the EWPT and that we can neglect the temperature dependence of $v_s$ and $m_s$. At first sight, one would expect loop diagrams involving the quartic self-coupling to have important effects at high temperatures, which would restore the $\mathbb{Z}_2$ symmetry for $T \gtrsim v_s$~\cite{Vaskonen:2016yiu}. However, since the light scalars are never in thermal equilibrium, these contributions are in fact absent and the only temperature dependence arises from interactions between the light scalars and SM Higgs bosons. It was shown in Ref.~\cite{Babu:2014pxa} that for $\lambda_{hs} > 0$ these interactions will restore the $\mathbb{Z}_2$ symmetry at high temperatures. Nevertheless, the dominant production mode for small $v_s$ is $h \to ss$, which does not require mixing between the two Higgs bosons. This production mechanism therefore works in exactly the same way also if the $\mathbb{Z}_2$ symmetry is still unbroken after the EWPT. The discussion below is therefore independent of whether the $\mathbb{Z}_2$ symmetry breaks before or after the EWPT.\footnote{We note that it was also shown in Ref.~\cite{Babu:2014pxa} that the energy density of domain walls resulting from the breaking of the $\mathbb{Z}_2$ is sufficiently small to be consistent with observational constraints.}

\section{Evolution of the dark sector}
\label{sec:evolution}

Once the number density of Higgs bosons becomes strongly Boltzmann suppressed, the freeze-in production of light scalars terminates and the two sectors of the theory develop completely independently. This does, however, not necessarily mean that the co-moving number density of light scalars stays constant. Since the $\mathbb{Z}_2$ symmetry of the light scalar is broken, $2 \to 3$ and $3 \to 2$ processes can change the number density and temperature of the dark sector. In particular, these processes can lead to a period of chemical equilibrium within the dark sector, which ends when the number-changing processes freeze out. In the following we will first determine the regions of parameter space where interactions within the dark sector can be important and then develop the necessary formalism for calculating the final abundance of light scalars in these regions.

\subsection{Step 4: Thermalisation}

All the production mechanisms discussed above lead to the production of light scalars with substantial energy. For example, scalars produced in Higgs decays will have an energy of $m_h / 2 \gg m_s$. Scattering between light scalars will bring them into kinetic equilibrium with each other, such that we can define a dark sector temperature $T_\text{dark}$, which will in general be different from the temperature $T_\text{SM}$ of the SM particles.\footnote{In principle, scattering between light scalars and SM fermions via Higgs exchange could lead to the exchange of energy and hence kinetic equilibrium between the two sectors. The cross section for this process is parametrically given by $\sigma_\text{scat} \propto \lambda_{hs}^2 \, m_f^2 / m_h^4$. The corresponding scattering rate $\Gamma_\text{scat} = \sigma_\text{scat} n_f$ is found to be tiny compared to the Hubble rate for the range of $\lambda_{hs}$ that we are interested in, so it is safe to neglect these processes.} It is however not clear a priori whether the light scalars will also achieve chemical equilibrium, such that their distribution would follow an equilibrium distribution with vanishing chemical potential $\mu_\text{dark}$.\footnote{Note that chemical equilibrium does in general not imply that the chemical potentials vanish, but that they add up to zero. In our case, however, we only consider number-changing processes with final and initial states of the same particle species, so chemical equilibrium does imply vanishing chemical potential.} Indeed, we will now show that it is inconsistent to assume $\mu_\text{dark} = 0$ at the end of freeze-in.

In analogy to the co-moving number density $Y_s$ we define a rescaled energy density
\begin{equation}
Z_s = \frac{x_\mathrm{SM} \, \rho_s}{s_\mathrm{SM} \, m_s} \; .
\end{equation}
For a completely decoupled relativistic dark sector, $\rho_s \propto T_\mathrm{SM} \, n_s$ and hence $Z_s \propto Y_s = \text{const}$. If the dark sector is populated through Higgs decays, on the other hand, each Higgs decay changes the total number of particles in the dark sector by 2 and the total energy by $m_h$. Thus $\mathrm{d} \rho_s = \tfrac{m_h}{2} \mathrm{d} n_s$, and the evolution of $Z_s$ is simply given by
\begin{equation}
\frac{\mathrm{d}Z_s}{\mathrm{d}x_\mathrm{SM}} = \frac{m_h}{2} \, x_\mathrm{SM} \frac{\mathrm{d}Y_s}{\mathrm{d}x_\mathrm{SM}} \; , 
\end{equation}
where $\mathrm{d}Y_s/\mathrm{d}x_\mathrm{SM}$ is given in eq.~(\ref{eq:Higgs_decay}). For $m_s \ll m_h$, we then find $Z_s/Y_s = 1.74$ at the end of freeze-in, which can be rewritten as
\begin{equation}
 \frac{\rho_s}{n_s} = 1.74 \, T_\mathrm{SM} \; .
 \label{eq:rhoovern}
\end{equation}
In other words, freeze-in production of light scalars imposes a specific ratio of energy density to number density in the dark sector.

Given kinetic equilibrium, the number density and energy density of light scalars after freeze-in can be written as
\begin{align}
 n_{s,\mathrm{fi}} & = \frac{1}{2\pi^2} \int \frac{k^2 \, \mathrm{d}k}{\exp\left[(E-\mu_\text{dark,fi})/T_\text{dark,fi}\right] - 1} \\
 \rho_{s,\mathrm{fi}} & = \frac{1}{2\pi^2} \int \frac{E \, k^2 \, \mathrm{d}k}{\exp\left[(E-\mu_\text{dark,fi})/T_\text{dark,fi}\right] - 1} \; ,
\end{align}
where $T_\text{dark,fi}$ and $\mu_\text{dark,fi}$ denote the temperature and chemical potential of the dark sector at the end of freeze-in. The precise definition of when freeze-in ends is of course arbitrary and does not impact the subsequent calculation. We can now use eq.~(\ref{eq:rhoovern}) to eliminate $T_\text{dark,fi}$. For example, if we neglect the $-1$ in the denominator of the expressions above and assume that the light scalars are highly relativistic, it can be immediately seen that $\rho_{s,\mathrm{fi}} / n_{s,\mathrm{fi}} = 3 \, T_\text{dark,fi}$ independent of $\mu_\text{dark,fi}$. In combination with eq.~(\ref{eq:rhoovern}) we therefore find $T_\text{dark,fi} = 0.58 \, T_\text{SM,fi}$. In this approximation, the temperature of the dark sector is independent of the \emph{amount} of freeze-in production, which can therefore only be captured by choosing an appropriate chemical potential $\mu_\text{dark}$. 

It should be clear from the discussion above that it will in general only be possible to match the freeze-in prediction for both $n_s$ and $\rho_s$ if both $T_\text{dark,fi}$ and $\mu_\text{dark,fi}$ are allowed to vary.\footnote{Refs.~\cite{Chu:2011be,Heikinheimo:2016yds} suggest to calculate the dark sector temperature by using the Stefan-Boltzmann law $T_\mathrm{dark} = \left[\rho_s \, g^\ast / (\rho \, g_\mathrm{dark}^\ast) \right]^{1/4} T_\mathrm{SM}$, which implicitly assumes vanishing chemical potential. While this formula gives the correct prediction for the energy density, it fails to predict the correct number density in the dark sector and therefore cannot be applied to our case.}
This implies in particular that the number density of light scalars may initially be quite different from the equilibrium number density expected for the temperature $T_\text{dark}$. Whether or not it stays that way, i.e.\ whether or not the light scalars enter into chemical equilibrium, depends on the rate of number-changing processes. 

Assuming for the moment that these processes are inefficient, the subsequent evolution of the dark sector will be fully determined by the simultaneous conservation of the dark sector entropy and co-moving number density. The initial entropy density of the dark sector is given by
\begin{equation}
 s_\text{dark,fi} = \frac{\rho_{s,\mathrm{fi}} + p_{s,\mathrm{fi}}- \mu_\text{dark,fi} \, n_{s,\mathrm{fi}}}{T_\text{dark,fi}}  \; , 
\end{equation}
while the co-moving number density is simply given by $Y_s = n_{s,\mathrm{fi}} / s_\text{SM,fi}$, because the contribution of the dark sector to the Hubble expansion rate is completely negligible. Since entropy is separately conserved in the two sectors, the ratio of the entropy densities
\begin{equation}
 \xi = \frac{s_\text{SM,fi}}{s_\text{dark,fi}} \; ,
\end{equation}
is simply a constant.

We can therefore calculate the temperature and chemical potential of the dark sector, parametrised by $x_\text{dark} = m_s / T_\text{dark}$ and $\nu_\text{dark} = \mu_\text{dark} / T_\text{dark}$ as a function of the visible sector temperature $T_\text{SM}$. In the absence of number-changing processes, the evolution of these quantities is fairly trivial. The only effect that changes the temperature ratio and the chemical potential is when particles in either of the two sectors become non-relativistic.

\begin{figure}
\centering
	\includegraphics[scale=0.7,clip,trim = 20 80 40 70]{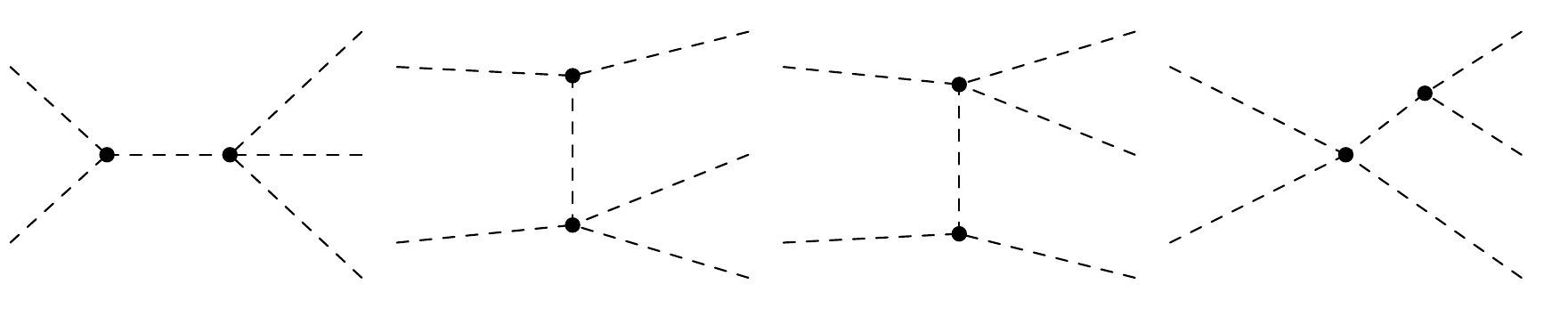}
\caption{\label{fig:2to3}Number-changing processes in the dark sector which result in a decrease in the dark sector temperature and an increase in the dark matter number density.}
\end{figure}

We can use these observations to answer the question whether it is consistent to ignore the effect of number-changing processes. The most important such processes are the $2 \to 3$ processes shown in figure~\ref{fig:2to3}, which convert kinetic energy into new particles, such that the dark sector temperature decreases while the co-moving number density increases. For $m_s \ll \sqrt{S}$ the cross section for the simplest of these processes (shown in the leftmost panel) is given by\footnote{The cross section is in fact proportional to $\lambda_s^4 \, v_s^2$, which we write as $2 \, \lambda_s^3 \, m_s^2$.}
\begin{equation}
\sigma_{2 \to 3} = \frac{27 \, \lambda_s^3 m_s^2}{64 \pi^3 S^2} \; .
\end{equation}
We find however that the additional diagrams give a larger contribution to the total cross section, because the intermediate particles can be nearly on-shell. For these diagrams it becomes unfeasible to perform the phase-space integration analytically (in particular in the important limit $\sqrt{S} \to 3\,m_s$), so that we use \texttt{calchep 3.7.1}~\cite{Belyaev:2012qa} to obtain the full result for arbitrary centre-of-mass energies.

The corresponding interaction rate is given by $\Gamma_{2 \to 3} = \langle \sigma v\rangle_{2 \to 3} \, n_s$, where $\langle \sigma v \rangle_{2\to3}$ denotes the thermal average \emph{with respect to the temperature of the dark sector}, which can be calculated in terms of the SM temperature. If $\Gamma_{2 \to 3}(T_\text{SM}) \ll H(T_\text{SM})$ for all temperatures $T_\text{SM}$, the light scalars will never enter into chemical equilibrium with each other. In this case, we can indeed treat the co-moving number density $Y_s$ as constant and obtain the final relic abundance of light scalars directly from the freeze-in yield.

\begin{figure}
\centering
	\includegraphics[width=.925\textwidth]{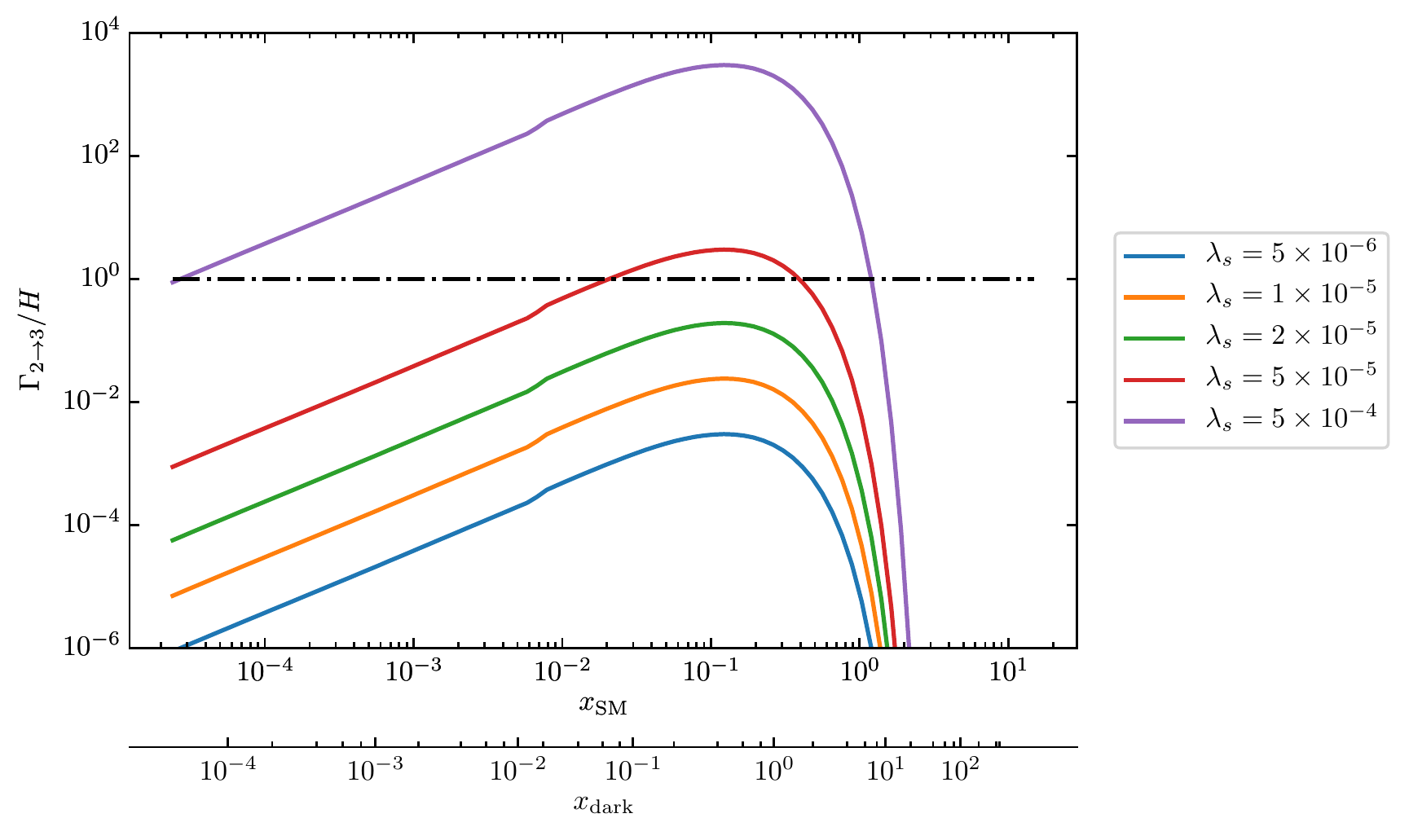}
\caption{Ratio of the rate of number-changing processes, $\Gamma_{2\to3} = \langle \sigma_{2 \to 3}\rangle \, n_s$ and the Hubble expansion rate $H$ as a function of the inverse SM temperature $x_\mathrm{SM}$. We consider fixed $m_s = 1\,\mathrm{MeV}$ and several different values of $\lambda_s$. A second $x$-axis indicates the corresponding inverse temperature of the dark sector $x_\text{dark}$ under the assumption that number-changing processes are negligible. Initially, this temperature is given by $x_\text{dark} \approx 1.72\, x_\mathrm{SM}$ but then evolves independently due to the separate conservation of entropy in the two sectors.\label{fig:rates}}
\end{figure}

We show in figure~\ref{fig:rates} the ratio $\Gamma_{2 \to 3} / H$ for specific values of $\lambda_s$ and $m_s$ under the assumption that the co-moving number density is given by $Y_s = \Omega_\text{DM} \, \rho_\text{c,0} / (m_s \, s_\text{SM,0}) = \text{const}$, such that light scalars would constitute all of the DM in the present Universe. The two $x$-axis of the figure indicate $x_\text{SM}$ and $x_\text{dark}$, respectively. We observe that the interaction rate is largest relative to the Hubble rate for $x_\text{dark} \approx 0.4$, i.e.\ shortly before the average centre-of-mass energy becomes insufficient to produce a new particle. Since the cross section is simply proportional to $\lambda_s^3$ for fixed $m_s$, we can immediately infer the largest value of $\lambda_s$ for which $\Gamma_{2 \to 3} / H < 1$ for all temperatures. For example, for $m_s = 1\,\text{MeV}$ we find $\lambda_s \lesssim 3\cdot 10^{-5}$. 

\begin{figure}
\centering
	\includegraphics[width=.6\textwidth]{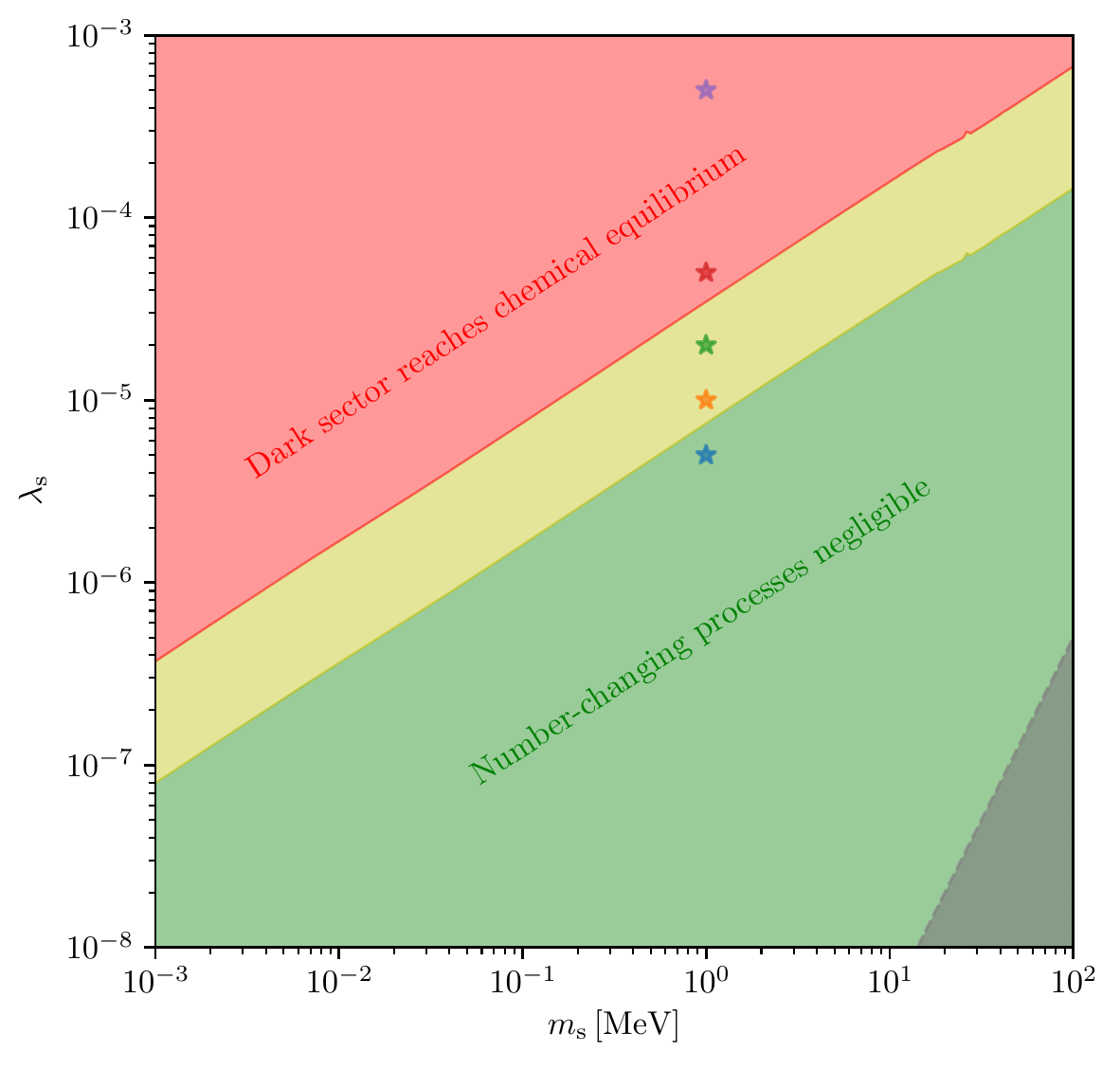}
\caption{\label{fig:paramspace}Regions of different dark sector evolution as a function of $m_s$ and $\lambda_s$. In the red shaded regions the reaction rate $\langle\sigma v\rangle_{2\to3}\, n_s $ exceeds the Hubble rate $H(T)$ at some point in the early Universe, leading to a thermalised dark sector with vanishing chemical potential. In the yellow shaded regions, the ratio of the rates is between 1\% and unity, such that number-changing processes cannot be neglected even though the dark sector might not reach chemical equilibrium. In the green shaded regions, the reaction rate never exceeds 1\% of the expansion rate, such that the abundance will not be significantly affected by number-changing processes. The grey area in the bottom-right corner indicate the parameter range where $2\to2$ processes give the dominant contribution to the freeze-in yield.}
\end{figure}

We illustrate in figure~\ref{fig:paramspace} how the maximum value of $\Gamma_{2 \to 3} / H$ depends on the scalar mass $ m_s $ and its self coupling $ \lambda_s $ under the assumption that $Y_s$ is constant and corresponds to the observed DM relic abundance. The five stars correspond to the five curves shown in figure~\ref{fig:rates}. In the red shaded parameter region the maximum ratio exceeds unity, meaning that $2\to3$ processes will be efficient enough to bring the dark sector into chemical equilibrium. In the green shaded region, on the other hand, the ratio never exceeds 0.01 and we can safely neglect number-changing processes. In the intermediate region, shaded in yellow, the reaction rate stays below the Hubble expansion rate, but is still large enough that we cannot treat $Y_s$ as constant. We will deal with this case together with the case of full chemical equilibrium in the following section.

Finally, we also indicate in figure~\ref{fig:paramspace} the parameter region where the freeze-in abundance is dominantly set by $t$-channel processes (rather than Higgs decays). We find that in these parameter regions $2\to3$ processes are completely negligible, so that the evolution of the dark sector after the end of freeze-in is trivial. Conversely, in the regions of parameter space where chemical equilibrium may be established, it is fully justified to estimate the energy density of the dark sector under the assumption of freeze-in via Higgs decays, as we have done above.

\subsection{Step 5: Dark sector freeze-out}

Let us now take a closer look at what happens when the dark sector reaches chemical equilibrium. In this case $\mu_\text{dark} = 0$ and the number density is given by the equilibrium distribution
\begin{equation}
 n^\text{eq}_{s} = \frac{1}{2\pi^2} \int \frac{k^2 \, \mathrm{d}k}{\exp\left(E/T_\text{dark}\right) - 1} \; .
\end{equation}
As the light scalars become non-relativistic, the equilibrium number density becomes exponentially suppressed. The reason is that the $2 \to 3$ processes, which helped to populate the dark sector, now become inefficient due to the lack of sufficient kinetic energy to produce a third particle. The inverse process, on the other hand, remains fully efficient and leads to a depletion of light scalars in combination with an increase of the dark sector temperature. This is the well-known cannibalism mechanism~\cite{Carlson:1992fn,Farina:2016llk,Pappadopulo:2016pkp}. It ceases to be efficient once the rate for the $3 \to 2$ process drops below the Hubble expansion rate, at which point the interactions freeze out and the co-moving number density of light scalars becomes constant. The final abundance of light scalars is then determined by the temperature when the departure from chemical equilibrium happens.

While it is possible to use entropy conservation to obtain an approximate estimate of the freeze-out abundance~\cite{Carlson:1992fn}, we will calculate the evolution of $Y_s$ numerically by solving the Boltzmann equation
\begin{equation}
 \frac{\mathrm{d}{Y_s}}{\mathrm{d}x_\text{SM}} = \frac{s_\text{SM}}{H\,x_\text{SM}} \langle\sigma v\rangle_{2\to3} Y_s^2 \left[1 - \frac{Y_s}{Y_s^\text{eq}}\right] \; ,
 \label{eq:Boltzmann_log}
\end{equation}
where $Y^\text{eq}_s = n^\text{eq}_s / s_\text{SM}$. To obtain this equation we have made use of the fact that
\begin{equation}
 \langle\sigma v\rangle_{2\to3} \, n^\text{eq}_s = \langle\sigma v^2\rangle_{3\to2} \,  (n^\text{eq}_s)^2 \; ,
 \label{eq:Boltzmann}
\end{equation}
because the rates of the two processes must be equal in thermal equilibrium (see appendix~A of Ref.~\cite{Bernal:2015bla}). The Boltzmann equation can now be solved for given $Y_{s,\mathrm{fi}}$ and $\langle\sigma v\rangle_{2\to3}$ to yield the present-day value of $Y_s$, $Y_{s,0}$. To do so, it is convenient to rewrite eq.~(\ref{eq:Boltzmann}) as 
\begin{equation}
 \frac{\mathrm{d}\log Y_s}{\mathrm{d}\log x_\text{SM}} = \frac{s_\text{SM}}{H} \langle\sigma v\rangle_{2\to3} Y_s \left[1 - \frac{Y_s}{Y^\text{eq}_s}\right] \; .
\end{equation}
Note that our approach differs from the one taken in Refs.~\cite{Bernal:2015bla,Bernal:2015ova} in that we base our calculation on $\langle\sigma v\rangle_{2\to3}$ rather than $\langle\sigma v^2\rangle_{3\to2}$, which we can calculate for arbitrary centre-of-mass energies (whereas Refs.~\cite{Bernal:2015bla,Bernal:2015ova} only consider the non-relativistic limit). This makes it possible to treat thermalisation and freeze-out in a unified framework, i.e.\ we can use the Boltzmann equation also in the case where the dark sector never reaches chemical equilibrium. 

The disadvantage of our approach is that we need to track the dark sector temperature throughout the evolution of the dark sector (whereas $\langle\sigma v^2\rangle_{3\to2}$ becomes independent of the dark sector temperature in the non-relativistic limit). To do so, we again make use of the fact that the entropy ratio $\xi$ remains constant, such that we can write
\begin{equation}
 \frac{1}{Y_s} = \frac{\xi \, s_\text{dark}}{n_s} = \xi \left(x_\text{dark} \frac{\rho_s + p_s}{m_s \, n_s} - \nu_\text{dark}\right) \; ,
 \label{eq:Ys}
\end{equation}
where as before $x_\text{dark} = m_s / T_\text{dark}$ and $\nu_\text{dark} = \mu_\text{dark} / T_\text{dark}$. If we neglect the $-1$ in the denominator of the Bose-Einstein distribution, which is a good approximation as long as the dark sector is sufficiently sparsely populated, we find that
\begin{equation}
\zeta \equiv \frac{\rho_s + p_s}{m_s \, n_s} 
\end{equation}
depends only on $x_\text{dark}$, but not on $\nu_\text{dark}$, which in turn is given by
\begin{equation}
 \nu_\text{dark} = \log \left(\frac{Y_s}{Y^\text{eq}_s}\right) \; .
\end{equation}
Combining these equations gives
\begin{equation}
 \frac{1}{Y_s} = \xi \left[x_\text{dark} \, \zeta(x_\text{dark}) - \log \left(\frac{Y_s}{Y^\text{eq}_s(x_\text{dark})}\right) \right] \; ,
 \label{eq:general_Y_from_xi}
\end{equation}
which is now independent of $\nu_\mathrm{dark}$. Eq.~(\ref{eq:general_Y_from_xi}) can be solved numerically for $x_\text{dark}$ for given $Y_s$, $\xi$ and $s_\text{SM}$ (the latter entering in the calculation of $Y^\text{eq}_s$). 

\begin{figure}
\centering
	\includegraphics[width=.925\textwidth]{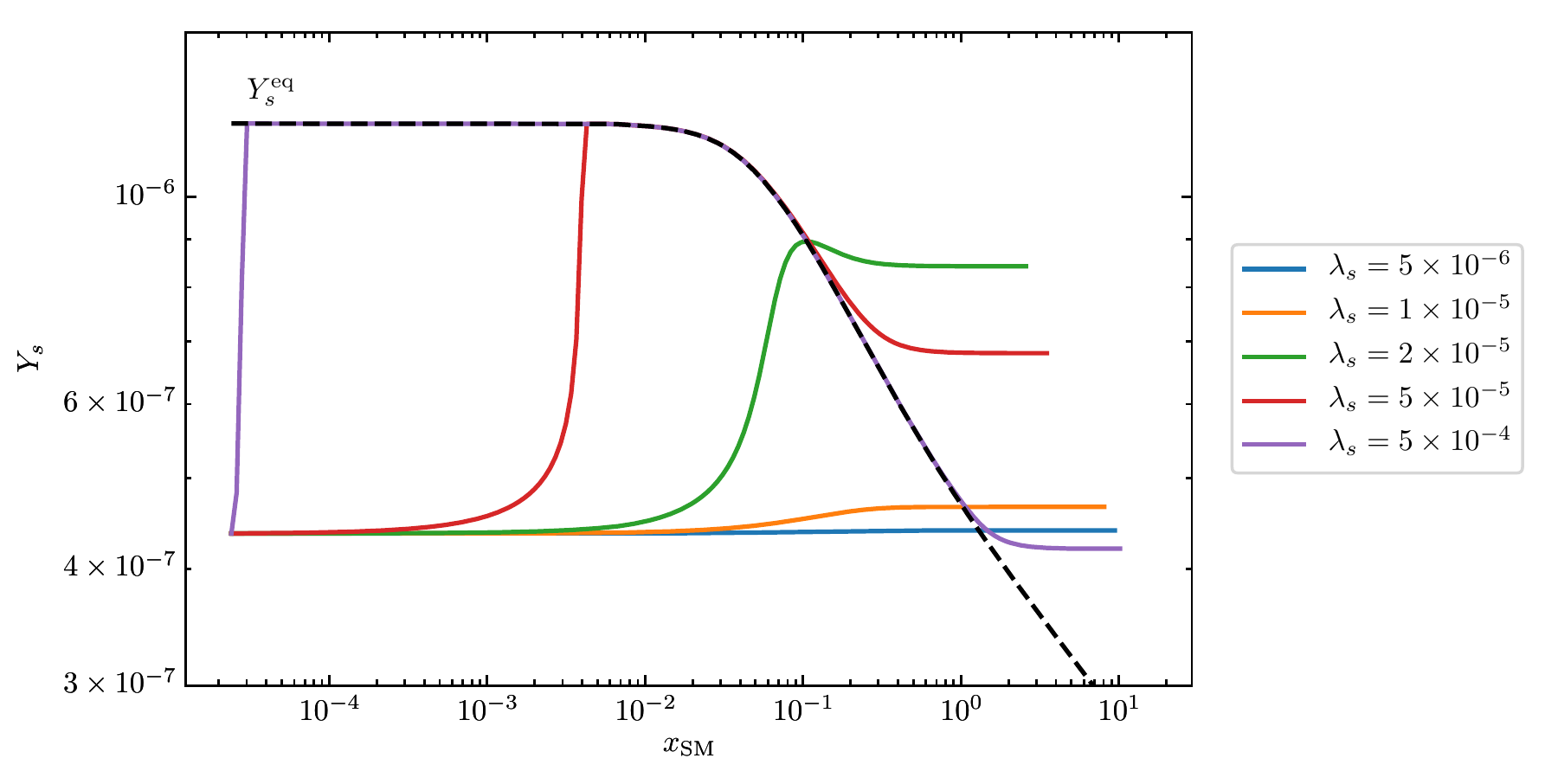}
\caption{\label{fig:Boltzmann} Evolution of $Y_s$ as a function of $x_\mathrm{SM}$ for fixed scalar mass $m_s = 1\,\text{MeV}$ and different values of the coupling $\lambda_s$. The initial value for $Y_s$ is chosen such that in the absence of number-changing processes the light scalars would constitute all of DM in the present Universe.}
\end{figure}

We are now in the position to solve eq.~(\ref{eq:Boltzmann}) numerically (see appendix~\ref{app:boltzmann} for details). Figure~\ref{fig:Boltzmann} shows a number of examples for how $Y_s$ evolves as function of $x_\text{SM}$. All examples correspond to the same scalar mass, $m_s = 1\,\mathrm{MeV}$, but different values of $\lambda_s$ (as indicated in figure~\ref{fig:paramspace}). Note that $x_\text{dark}$ depends on $\lambda_s$ and hence we do not show a second $x$-axis in this plot. We identify the following cases (cf.\ figure~\ref{fig:paramspace}):
\begin{itemize}
\item For $\lambda_s \lesssim 5 \cdot 10^{-6}$ the reaction rate always stays well below the Hubble rate and the co-moving number density does not change.
\item For $\lambda_s \sim 10^{-5}$ the reaction rate stays only slightly below the Hubble rate. The $2\to3$ processes can therefore not be entirely neglected and lead to a slight increase in $Y_s$. Nevertheless, the dark sector does not reach chemical equilibrium in this case.
\item For $\lambda_s \gtrsim 3 \cdot 10^{-5}$ the dark sector reaches chemical equilibrium and subsequently freezes out. With increasing $\lambda_s$ equilibrium is reached more quickly and is maintained longer before freeze-out.
\end{itemize}
As for standard freeze-out we therefore find that larger couplings lead to a smaller relic abundance. For the case of freeze-out via $3\to2$ processes, however, the dependence on the couplings is much milder. Indeed, as shown in appendix~\ref{app:boltzmann}, the co-moving number density after dark sector freeze-out can approximately be written as
\begin{equation}
 Y_s = \frac{4}{3 \, \xi \left[ \log\left(\frac{M_\mathrm{p} \, \lambda_s^3}{m_s \, \xi^{2/3}}\right) - a\right]}
\end{equation}
with $a \approx 4\text{--}5$. In particular, $Y_s$ depends only logarithmically on $\lambda_s$ and therefore changes only slightly even if $\lambda_s$ varies substantially.

\subsection{Relic abundance of light scalars}

\begin{figure}
\centering
\includegraphics[width=.47\textwidth]{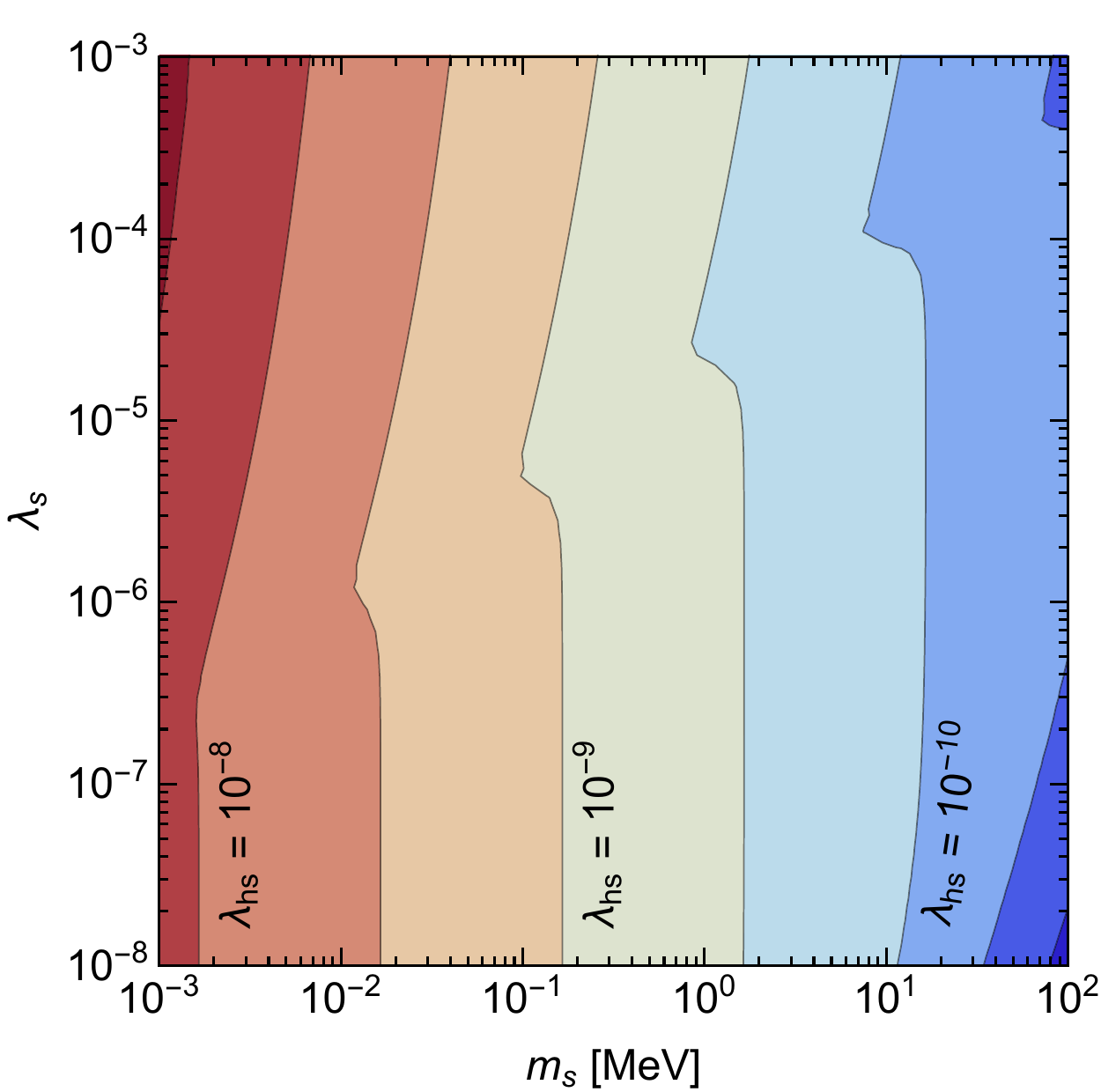}\hfill\includegraphics[width=.47\textwidth]{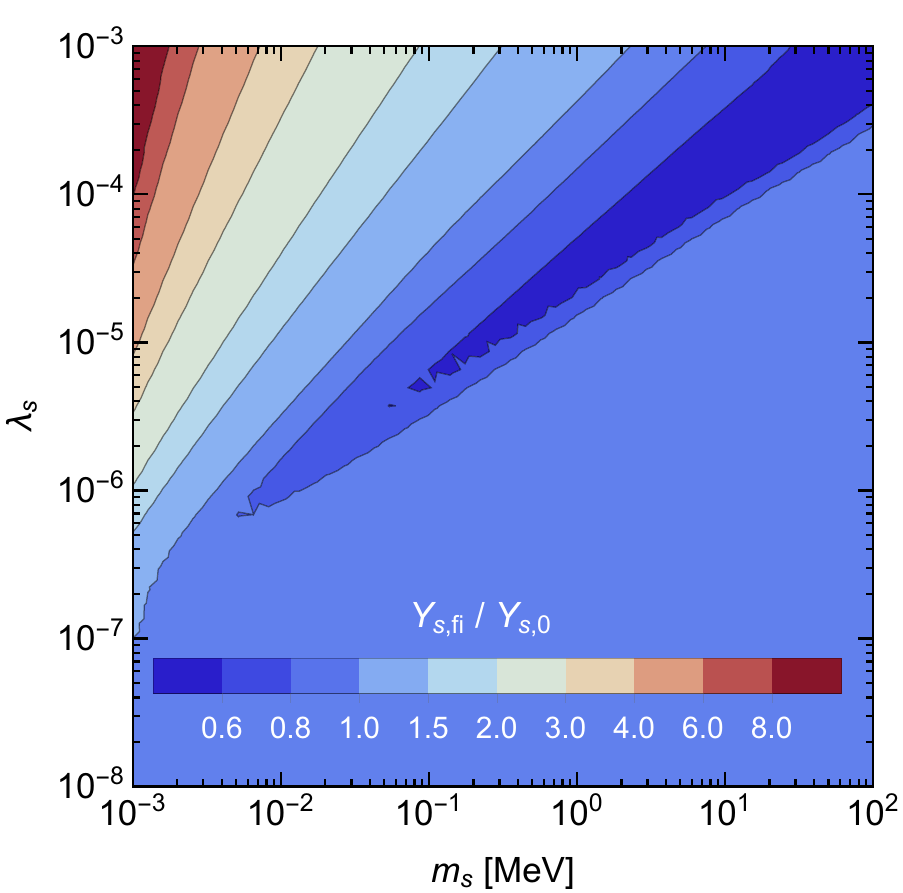}
\caption{\label{fig:relic} Left: Value of $\lambda_{hs}$ required to obtain the correct abundance from freeze-in, such that at the end of dark sector freeze-out the light scalars account for all of DM. Right: Ratio of the scalar abundance at the end of freeze-in, $Y_{s,\text{fi}}$, and the scalar abundance today, $Y_{s,0}$, under the assumption that light scalars constitute all of DM. One can clearly identify parameter regions where $2\to3$ processes lead to an increase in the abundance ($Y_{s,0} > Y_{s,\text{fi}}$) and regions where dark sector freeze-out leads to a depletion of light scalars ($Y_{s,0} < Y_{s,\text{fi}}$).}
\end{figure}

We now have all the necessary ingredients to combine the calculations outlined in steps 1--5 above and calculate the final relic abundance of light scalars as a function of $m_s$, $\lambda_s$ and $\lambda_{hs}$. We show in the left panel of figure~\ref{fig:relic} the value of $\lambda_{hs}$ needed to obtain $\Omega_s h^2 = 0.12$ as a function of $m_s$ and $\lambda_s$. Following eq.~(\ref{eq:lambdahs}), one would expect $\Omega_s$ to be independent of $\lambda_s$ in the absence of $2\to3$ processes, such that lines of constant $\lambda_{hs}$ would simply be vertical. This is indeed what we find for small values of $\lambda_s$, when number-changing processes are inefficient.

For larger self-couplings, however, the behaviour becomes more complicated and lines of constant $\lambda_{hs}$ are no longer vertical. This effect can be most easily understood by consulting the right panel of figure~\ref{fig:relic}, which shows the ratio of the scalar abundance after freeze-in, $Y_{s,\text{fi}}$, and the present-day abundance after dark sector freeze-out, $Y_{s,0}$, for parameter points that satisfy the relic density requirement. In the bottom-right corner we see once again that number-changing processes have a negligible effect on the abundance of light scalars, such that $Y_{s,\text{fi}} / Y_{s,0} = 1$. For somewhat larger values of $\lambda_s$ (indicated by dark blue shading) the dark sector still does not reach chemical equilibrium, but the effect of $2\to3$ processes cannot be completely neglected, leading to a net increase in the co-moving number density (and a decrease the dark sector temperature relative to the SM temperature). This means that $Y_{s,\text{fi}} / Y_{s,0} < 1$ and hence smaller freeze-in yields (and smaller values of $\lambda_{hs}$) are sufficient to reproduce the observed DM abundance in the present Universe. For even larger values of $\lambda_s$ (indicated by orange and red shading) the dark sector enters into chemical equilibrium and dark sector freeze-out leads to a decrease in the co-moving number density, as well as to heating of the dark sector. Since in this parameter region $Y_{s,\text{fi}} / Y_{s,0} > 1$, larger freeze-in yields (and hence larger values of $\lambda_{hs}$) are required to reproduce the observed DM relic abundance. 

Finally, we note that in the bottom-right corner of the left panel of figure~\ref{fig:relic} the freeze-in contribution of additional channels beyond Higgs decays becomes important (see section~\ref{sec:afterEWSB}), leading to an additional dependence of $Y_s$ on $\lambda_s$, such that lines of constant $\lambda_{hs}$ are no longer vertical, even though the effect of number-changing processes is negligible.

\section{Phenomenological consequences}
\label{sec:pheno}

Let us now answer the question whether the light scalars considered above can constitute all of the DM in the present Universe. For this purpose we need to consider their lifetime (which must not only exceed the age of the Universe but also satisfy various constraints from indirect detection) as well as the self-interaction cross section (which is constrained by a number of astrophysical observations). A similar discussion can be found in Ref.~\cite{Babu:2014pxa}.

\subsection{Light scalar decays}

In the present work we focus on scalar DM particles in the mass range between 1 keV and 100 MeV, such that there are only two possible decay modes: $s \to e^+e^-$ and $s \to \gamma\gamma$ (see also Ref.~\cite{Flacke:2016szy} for a similar discussion in the context of relaxion-Higgs mixing). The corresponding decay widths can be calculated in complete analogy to the case of a light Higgs boson~\cite{Gunion:1989we,Djouadi:2005gi}. The electronic partial decay width is given by
\begin{equation}
 \Gamma(s \to e^+e^-) = \sin^2 \theta \, \frac{m_s \, m_e^2 }{8\pi \, v^2} (1 - z_e)^{3/2}
\end{equation}
with $z_e = 4 m_e^2 / m_s^2$ and $\Gamma(s \to e^+e^-) = 0$ for $z_e \geq 1$. Decays into photons are absent at tree-level, but are induced via loops of SM fermions and bosons. The resulting partial width can be written as
\begin{equation}
  \Gamma(s \to \gamma\gamma) = \sin^2 \theta \, \frac{\alpha^2 \, m_s^3}{256\pi^3 \, v^2} \left| f(z_e) + \frac{7}{3}\right|^2
\end{equation}
where $\alpha$ is the fine-structure constant and 
\begin{equation}
 f(\tau) = 2 \tau\left[1+(1-\tau) \, \text{arctan}^2 \left(\frac{1}{\sqrt{\tau-1}}\right) \right]
\end{equation}
is the form factor for fermionic loops. We include this form factor only for electrons and evaluate all other loops in the limit $z_f \to \infty$, which is a good approximation even for up and down quarks, because their effective mass in the loop is of the order of the pion mass~\cite{Dolan:2014ska}.

As expected, we find that decays into electrons completely dominate once they become kinematically allowed. In fact, for the typical coupling strengths required for freeze-in production ($\sin \theta \sim 10^{-12}\text{--}10^{-14}$), the resulting lifetime is well below $10^{20}\,\mathrm{s}$. Such lifetimes are strongly excluded by observations of the Cosmic Microwave Background~\cite{Aghanim:2018eyx,Poulin:2016anj,Stocker:2018avm} and, more recently, measurements of the 21cm radiation temperature in the re-ionisation epoch~\cite{Bowman:2018yin,Clark:2018ghm}. In other words, only a small fraction of DM could be in the form of unstable MeV scalars produced via the freeze-in mechanism.

For $m_s < 1\,\mathrm{MeV}$ only the photonic decay mode remains kinematically allowed and the scalar lifetimes become much larger. At the same time, however, the observational signature is much more striking: a mono-energetic $\gamma$-ray or x-ray line with energy $E_\gamma = m_s / 2$. Such lines have been searched for in the context of keV sterile neutrinos~\cite{Adhikari:2016bei}. To convert a bound on the active-sterile mixing angle $\sin^2 2\theta$ into a bound on the scalar lifetime, we note that
\begin{equation}
\Gamma_N = \frac{9\,\alpha\,G_\mathrm{F}^2}{2048\pi^4}\sin^2 2\theta \, m_s^5 
\end{equation}
and that the bound on the scalar lifetime is stronger by a factor of 2, because each scalar decay produces two photons rather than one. The strongest bounds come from INTEGRAL~\cite{Yuksel:2007xh} for $m_s \sim 100\text{--}1000\,\mathrm{keV}$, from NuSTAR~\cite{Perez:2016tcq} for $m_s \sim 10\text{--}100\,\mathrm{keV}$ and from a combination of x-ray observations of M31~\cite{Horiuchi:2013noa} for even smaller masses. These constraints typically require $\tau \gtrsim 10^{27}\text{--}10^{29} \, \mathrm{s}$.

\subsection{Self-interactions and structure formation}

The quartic self-coupling between the light scalars gives rise to velocity-independent self-interactions. The corresponding cross section is given by~\cite{Heikinheimo:2016yds}
\begin{equation}
 \frac{\sigma_s}{m_s} = \frac{9\,\lambda_s^2}{32\pi \, m_s^3} \; .
\end{equation}
Observations from the Bullet Cluster constrain this cross section to be smaller than approximately $1\,\mathrm{cm^2\,g^{-1}}$~\cite{Markevitch:2003at,Randall:2007ph,Kahlhoefer:2013dca}, which translates to
\begin{equation}
 \lambda_s \lesssim 0.007 \left(\frac{m_s}{1\,\mathrm{MeV}}\right)^{3/2} \; .
\end{equation}

Since we are considering DM particles with mass in the keV range, another potential concern are constraints from structure formation, in particular from the Lyman-$\alpha$ forest~\cite{Murgia:2017lwo,Kobayashi:2017jcf}. The impact of our model on structure formation is however very different from the case of sterile neutrinos. The reason is that self-interactions between DM particles prevent free-streaming, and structures can therefore only be washed out by diffusion.\footnote{Different ways to evade structure formation constraints for keV-scale DM are discussed in Refs.~\cite{Heeck:2017xbu,Hansen:2017rxr}.} The diffusion length $l_s$ in the presence of self-interactions is given by~\cite{Chu:2015ipa}
\begin{equation}
 l_s^2 = \int_0^{t_\mathrm{dec}} \frac{\mathrm{d}t \, \langle v_s \rangle^2}{a^2 \, n_s \, \langle \sigma_s v_s \rangle} \; .
\end{equation}
The integration range in principle extends to the time $t_\mathrm{dec}$ when self-interactions become inefficient, i.e. when $n_s \langle \sigma_s v_s \rangle < H$, but the main contribution to the diffusion stems from relativistic particles. We can therefore approximate
\begin{equation}
 l_s^2 \approx \int_0^{a_\mathrm{NR}} \frac{\mathrm{d}a}{H \, a^3 \, n_s \, \langle \sigma_s v_s \rangle} \; ,
 \label{eq:ls}
\end{equation}
where $a_\mathrm{NR}$ denotes the scale factor when $x_\mathrm{dark} \sim 1$. 

If self-interactions (and hence $3\to2$ processes) are extremely weak, we can treat the temperature ratio $\alpha \equiv T_\mathrm{dark} / T_\mathrm{SM}$ as constant, with values typically in the range $\alpha \approx 0.1\text{--}0.3$. Using $n_s = Y_0 \, s_\mathrm{SM}$ and
\begin{equation}
\langle \sigma_s v_s \rangle = \frac{9 \, \lambda_s^2}{64 \pi \, T_\mathrm{dark}^2} 
\end{equation}
for $T_\mathrm{dark} \gg m_s$, it is easy to see that the integrand of Eq.~(\ref{eq:ls}) is independent of $a$ during radiation domination (ignoring the slight temperature dependence of $g_\mathrm{SM}^\ast$) and hence
\begin{equation}
 l_s^2 \propto \frac{a_\mathrm{NR} \, \alpha^2}{Y_0 \, \lambda_s^2} \; .
\end{equation}

Using the same approximation as above, $a_\mathrm{NR}$ is defined by $\alpha T_\mathrm{SM} = m_s$ and hence $a_\mathrm{NR} \propto \alpha / m_s$. Since $Y_0 \propto 1/m_s$, we therefore find that $l_s$ is largely independent of $m_s$. This leads to the extremely simple result
\begin{equation}
 l_s \sim 10^{-11}\,\mathrm{Mpc} \frac{\alpha^{3/2}}{\lambda_s} \; .
\end{equation}
Hence, for $\lambda_s > 10^{-11}$ the matter power spectrum remains unaffected on observable scales.

We note however that $a_\text{dec}$ can potentially be significantly larger than $a_\text{NR}$, such that dark matter behaves like a non-relativistic collisional fluid until rather late times. For example, for $m_s = 10\,\mathrm{keV}$ and $\lambda_s = 10^{-6}$ we find $a_\text{dec} \approx 2 \times 10^{-5}$ (corresponding to $T_\text{SM,dec} \approx 10\,\mathrm{eV}$). Such large values of $a_\text{dec}$ may affect the growth of density perturbations and hence potentially lead to modifications of the CMB~\cite{Das:2018ons}. We leave a study of these effects, which may strengthen the bound on the self-interaction cross section, to future work.

\subsection{Results}

\begin{figure}
\centering
	\includegraphics[width=.5\textwidth]{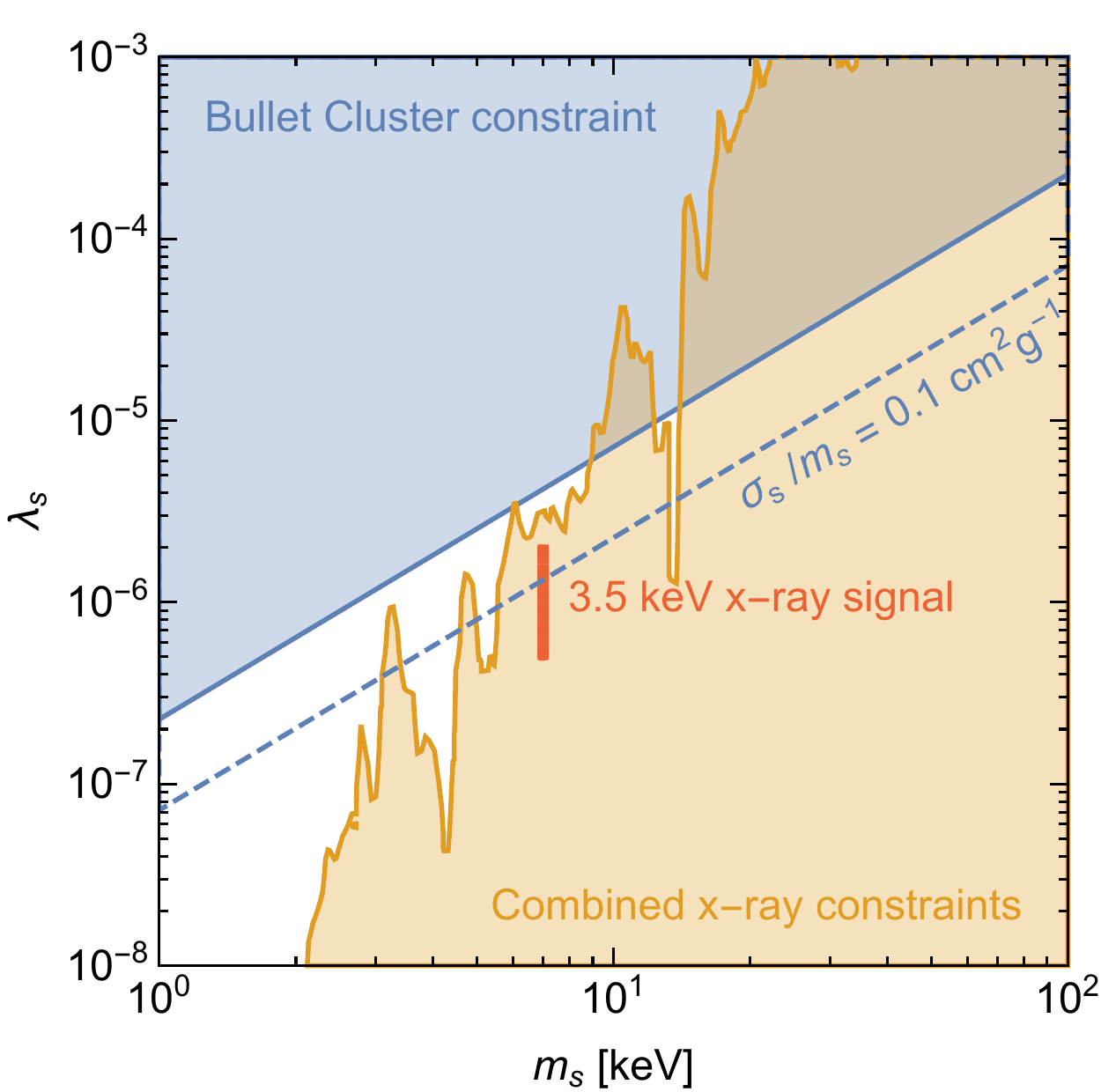}
\caption{\label{fig:results} Constraints on decaying lights scalars from a combination x-ray observations (orange) and from the bound on self-interactions obtained from the Bullet Cluster (blue). The dashed blue line indicates a self-interaction cross section one order of magnitude below current bounds, while the region shaded in red corresponds to the parameter region preferred by the claimed excesses of x-ray emission around 3.5 keV.}
\end{figure}

The results of our analysis are summarised in figure~\ref{fig:results}. For each value of $\lambda_s$ and $m_s$ we fix $\lambda_{hs}$ such that the light scalars constitute all of the DM in the present Universe (see figure~\ref{fig:relic}). Across the entire plot the lifetime of the scalars exceeds the age of the Universe by many orders of magnitude, but for large values of $m_s$, as well as for small values of $\lambda_s$ (corresponding to large values of $v_s$ and hence to larger mixing with the SM Higgs boson) there are strong constraints from searches for x-ray lines. For small masses and large self-couplings, on the other hand, constraints from DM self-interactions become strong and exclude a large part of the parameter space. The two constraints are highly complementary and in combination rule out light scalars with masses above about 10 keV. For smaller scalar masses, however, viable parameter regions remain.

This mass range is of particular interest in the context of interpreting a number of claimed observations of an anomalous x-ray emission around 3.5 keV in several galaxy clusters~\cite{Bulbul:2014sua,Boyarsky:2014jta}. We indicate the parameter region where these excesses can be accommodated by a red box, noting the well-known tension between these claims and the constraints obtained from M31~\cite{Abazajian:2017tcc}. Intriguingly, this parameter region corresponds to self-interaction cross sections of the order of $0.1\,\mathrm{cm^2\,g^{-1}}$ (indicated by the dashed blue line), which are only an order of magnitude below current bounds and may be testable with novel approaches, such as measurements of core sizes in galaxy clusters~\cite{Kaplinghat:2015aga,Tulin:2017ara}.

In conclusion, should the x-ray line emission around 3.5 keV be confirmed by future observations, light scalars produced via the freeze-in mechanism may provide an attractive explanation. In contrast to sterile neutrinos, these particles do not free-stream in the early Universe, but the effects of self-interactions may play an important role. Light scalars therefore have a distinct phenomenology that can be explored with an improved understanding of structure formation at small scales.

\section{Conclusions}

In this paper we have studied in detail the cosmological evolution and phenomenological properties of light scalar bosons in the keV to MeV range that mix with the SM Higgs boson. While the interactions with the Higgs render the light scalars unstable, for sufficiently small couplings their lifetime can be large compared to the age of the Universe. In these parameter regions light scalars never enter into thermal equilibrium with SM particles, so that their abundance is set via the freeze-in mechanism. Nevertheless, self-interactions and number-changing processes may be sufficiently large that the light scalars thermalise amongst themselves, leading to a freeze-out in the dark sector.

A close investigation of the freeze-in production of light scalars reveals a number of subtleties not fully appreciated previously. First of all, we point out the importance of the electroweak phase transition, which dramatically changes the production channels and~-- crucially~-- renders the freeze-in yield insensitive to the reheating temperature. Furthermore, we discuss the effects of a temperature-dependent Higgs mass and vev, which can potentially lead to the production of light scalars through oscillations (although the resulting yield is found to be sub-dominant in practice). Finally, we have studied in detail the impact of a proper treatment of quantum statistics for the particles in the thermal bath and find substantial differences to the case of classical statistics.

Due to the existence of number-changing processes, the evolution of the dark sector does not terminate at the end of freeze-in. We calculate the rate of $2\to3$ processes in order to determine whether the dark sector reaches chemical equilibrium. In large regions of parameter space these processes play an important role and we solve the corresponding Boltzmann equation in order to calculate the abundance of light scalars after dark sector freeze-out. For the first time we consistently include the evolution of the dark sector temperature and chemical potential in the Boltzmann equation in order to obtain accurate results even when the dark sector does not fully thermalise or when particles are still semi-relativistic during dark sector freeze-out.

We find that it is possible for light scalars in the keV to MeV range to account for all of DM, provided the Higgs portal coupling $\lambda_{hs}$ is of the order of $10^{-10}\text{--}10^{-8}$. However, the combined constraints from the CMB, 21cm astronomy, x-ray observations and bounds on the self-scattering cross section require such light scalar DM particles to have a mass of less than about 10 keV (see figure~\ref{fig:results}). Excitingly, this is precisely the mass range where light scalars can provide an explanation of the anomalous x-ray emission at 3.5 keV observed in various astrophysical systems. In contrast to sterile neutrinos, light scalars do not free-stream and therefore do not suppress structures on small scales, but their self-interactions may lead to other observable effects such as the formation of cores in galaxy clusters.

Finally, we note that large regions of parameter space opens up if we do not require light scalars to constitute all of DM. In particular, there are effectively no bounds from DM self-interactions on a sub-dominant fraction of light scalars. Moreover, it would be interesting to explore the case where the light scalars have a lifetime that is small compared to the present Universe, such that they can modify the reionisation history. Such light scalars could for example act as the mediator between the SM and another stable DM particle, which also obtains its abundance via the freeze-in mechanism. A detailed study of these possibilities will be left for future work.

\acknowledgments

We thank Camilo Garcia-Cely, Sebastian Wild and Bryan Zaldivar for valuable comments on the manuscript, Maria Archidiacono, Basudeb Dasgupta, Andreas Goudelis, Jan Heisig, Julien Lesgourgues, Kai Schmidt-Hoberg and Tommi Tenkanen for enlightening discussions and the authors of \texttt{micrOmegas} for technical support. This work is funded by the DFG Emmy Noether Grant No.\ KA 4662/1-1. Numerical calculations were performed with computing resources granted by RWTH Aachen University under project rwth0280.

\appendix

\section{Solving the Boltzmann equation}
\label{app:boltzmann}

In this appendix we discuss how to solve the Boltzmann equation for $2\to3$ processes both numerically and analytically, and provide details on the implementation that we have employed to obtain our results.

Unlike in the case of standard thermal production of dark matter, where we can identify $ x_\mathrm{SM} = x_\mathrm{dark}$, the inverse temperature parameter of a decoupled dark sector $ x_\mathrm{dark} $ has a non-trivial dependence on $ x_\mathrm{SM} $, which is given by eq.~\eqref{eq:general_Y_from_xi}. Since in our set-up the light scalars are relativistic after freeze-in, $ x_\mathrm{dark} $ can only be determined by solving this equation numerically.\footnote{We note in passing that in the non-relativistic limit eq.~\eqref{eq:general_Y_from_xi} can be solved for $x_\mathrm{dark}$, giving 
\[x_\mathrm{dark}=\frac{1}{2\pi} \, \exp{\left(\frac{5}{3} - \frac{2}{\xi Y_s}\right)} \, m_{s}^{2} \, \left(Y_s s_\text{SM}\right)^{-\frac{2}{3}}.\]
We find, however, that this approximation is not sufficiently accurate for our purposes and therefore always use the exact numerical solution.} In our numerical implementation to solve the Boltzmann equation for $ 2\to3 $ processes, this step turns out to be the bottleneck.

Furthermore the structure of eq.~\eqref{eq:Boltzmann_log} corresponds to a so-called ``stiff'' differential equation. For this class of differential equations ODE-solving-algorithms without adaptive step size will fail to obtain the correct solution. It is therefore preferable to use algorithms that adapt their step size depending on how quickly the differential equation changes. For points where the right-hand side of the differential equation changes rapidly if the input parameters are varied, the chosen step size will be smaller than for points where the right-hand side of the differential equation is nearly invariant. To choose the appropriate step size, such adaptive algorithms need to evaluate the Jacobian of the right hand side of the differential equation, which in our case requires substantial computational expense.\footnote{Note that both $\langle \sigma v \rangle_{2\to3}$ and $Y_s^\text{eq}$ depend implicitly on $x_\text{dark}$, which in turn depends on $Y_s$ via the numerical solution of eq.~\eqref{eq:general_Y_from_xi}. It is hence necessary to determine the Jacobian numerically, which requires repeated calls of the function to determine $ x_\mathrm{dark} $ for given inputs $ \log{x_\mathrm{SM}} $ and $ \log{Y_s} $.} We therefore need to find a good trade-off between stability of the solution and the accuracy of the step size.

In our analysis we solve eq.~\eqref{eq:Boltzmann_log} using the implementation of the multistep-backward-differentiation method of the \texttt{GSL-v1.15}-library~\cite{Galassi:2011:GSL:1:15} which we found to give robust results with relatively few computational expensive steps. 

An important simplification is possible whenever we are not interested in tracking the full solution of the Boltzmann equation, but only in finding the final value of $ Y_s $. Indeed, if the coupling is large enough to thermalise the dark sector, it is usually not important when exactly chemical equilibrium is reached but it suffices to know that $ Y_s $ follows $ Y_s^\mathrm{eq} $ until the number-changing processes decouple. Indeed, such large couplings are computationally particularly expensive, so it is desireable to have an improved treatment. For this we adapt the method implemented in \texttt{MadDM}~\cite{Backovic:2013dpa} for the case of thermal freeze-out. Whenever it is safe to assume that chemical equilibrium is reached at some point, we make an initial guess of the value for $ x_\mathrm{SM,start} $ when number-changing processes decouple ($ \Gamma_{2\to3} \lesssim H $). We then start solving the Boltzmann equation at this point assuming $Y_s(x_\mathrm{SM,start}) = Y_s^\text{eq}(x_\mathrm{SM,start})$ and evolve to the regime where $ Y_s $ approaches a constant value. To determine the quality of our initial guess of $x_\mathrm{SM,start}$, we then move $ x_\mathrm{SM,start} $ to a smaller value (i.e.\ to higher temperatures) and repeat the evolution. This procedure is repeated until the final value of $ Y_s $ converges. This way we avoid solving the Boltzmann equation in the regime where $ \Gamma_{2\to3} \gg H $ and hence tiny step sizes would be needed for an accurate solution.

\begin{figure}
\centering
\includegraphics[width=.8\textwidth]{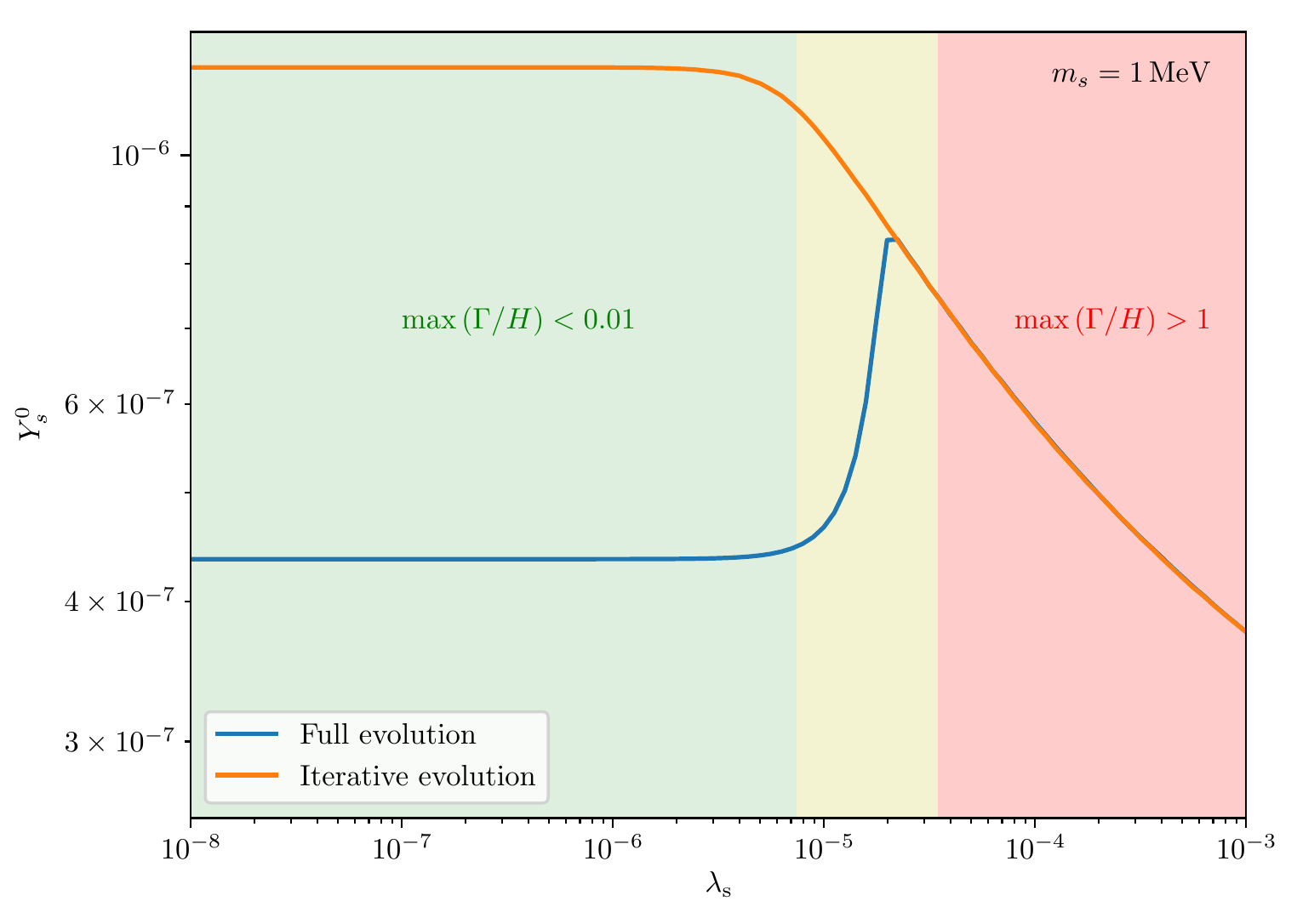}
\caption{\label{fig:BoltzmannSolver_comp} Comparison of the final result $ Y_{s}^{0} $ of both evolution methods for fixed mass as a function of the dark scalar self-coupling. The colour code of the background is the same as in figure~\ref{fig:paramspace}. For large couplings (when $ \max\left(\Gamma/H\right) \gtrsim 1 $) both methods agree, whereas for small couplings the method based on the assumption of chemical equilibrium leads results that are too large.}
\end{figure}

In figure \ref{fig:BoltzmannSolver_comp} we compare the results obtained from both methods for a fixed mass of $ m_s = 1\,\mathrm{MeV} $ as a function of the coupling $ \lambda_s $. We use the same background colours as in figure \ref{fig:paramspace} to indicate the naive ratio of $\Gamma_{2\to3} / H$. We see that both methods lead to the same results when the coupling is high enough (in the red regime) such that chemical equilibrium is maintained. For small couplings (green and yellow regimes) the results differ, as the assumption of chemical equilibrium, which is essential for the iterative method, is no longer satisfied. In this figure we can also see that the transition from the yellow to the red regime serves as a smooth transition between the two methods. In all our calcualtions we therefore use the iterative method for parameter point in the red regime and the full evolution for points in the yellow and green regime.

\bigskip

Finally, we note that under the assumption that DM is sufficiently non-relativistic during dark sector freeze-out, it is also possible to derive an approximate analytical estimate of the relic abundance after dark sector freeze-out. As always the condition for freeze-out is that
\begin{equation}
 H = n_{s,\mathrm{fo}} \, \langle \sigma v \rangle_{2\to3} = Y_{s,\mathrm{fo}} \, s_\mathrm{SM} \, \langle \sigma v \rangle_{2\to3} \; .
 \label{eq:Hf}
\end{equation}
In addition to the quantity of interest, i.e.\ the co-moving number density $Y_{s,\mathrm{fo}} = Y_{s,0}$ at freeze-out, this equation also depends on two further unknowns. The temperature of the visible sector at freeze-out, $T_\mathrm{SM,fo}$, enters via the Hubble rate and the SM entropy density, and the freeze-out temperature of the dark sector, $T_\mathrm{dark,fo}$, enters through the thermally averaged cross section. Indeed, in the non-relativistic regime, this cross section be approximately written as
\begin{equation}
 \langle \sigma v \rangle_{2\to3} \propto e^{-x_\mathrm{dark}} \frac{\lambda_s^3}{m_s^2} \; ,
\end{equation}
where the exponential reflects the fraction of DM particles with enough energy to induce a $2\to3$ process and the constant of proportionality needs to be determined numerically.

To eliminate $T_\mathrm{dark,fo}$, we can make use of the fact that in chemical equilibrium and for $x_\mathrm{dark} \gg 1$ eq.~(\ref{eq:Ys}) simplifies to~\cite{Carlson:1992fn}
\begin{equation}
 Y_s = \frac{1}{\xi \, x_\mathrm{dark}} \; .
 \label{eq:Ysnonrel}
\end{equation}
This leads to the somewhat surprising result that the right-hand side of eq.~(\ref{eq:Hf}) depends not linearly but exponentially on $Y_s$, in striking contrast to the case of standard freeze-out. As a result, we expect the DM abundance when freeze-out happens to depend only logarithmically on the fundamental model parameters. To make this explicit, we need to eliminate $T_\mathrm{SM,fo}$ by making use of the fact that $\xi = \mathrm{const}$, which implies
\begin{equation}
 x_\mathrm{SM}^3 \propto e^{x_\mathrm{dark}} \frac{g^\ast_\mathrm{SM} \, \sqrt{x_\mathrm{dark}}}{\xi}
\end{equation}
during radiation domination. 

Combining the above equations, one finds the approximate solution
\begin{equation}
 Y_s = \frac{4}{3 \, \xi \left[ \log\left(\frac{M_\mathrm{p} \, \lambda_s^3}{m_s \, \xi^{2/3}}\right) - a\right]} \; ,
 \label{eq:Ysapprox}
\end{equation}
where the parameter $a$ includes all the numerical factors not explictly included in the first term. By fitting to our results, we determine $a \approx 4\text{--}5$ with a slight dependence on both $x_\text{dark}$ and $x_\text{SM}$. This expression is found to give a good fit to the numerical solution of the Boltzmann equation provided that $\lambda_s$ is large enough for the non-relativistic approximation to be justified.\footnote{By comparing eq.~(\ref{eq:Ysnonrel}) to eq.~(\ref{eq:Ysapprox}) one can see that the non-relativistic limit is a good approximation if \[ \frac{3}{4}\left[ \log\left(\frac{M_\mathrm{p} \, \lambda_s^3}{m_s \, \xi^{2/3}}\right) - a\right] \gg 1 \; .\]} As anticipated, $Y_s$ depends only mildly on $\lambda_s$, such that varying $\lambda_s$ by orders of magnitude typically only changes $Y_s$ by a factor of a few. Finally, note that the freeze-in yield enters only via the entropy ratio $\xi$. Since larger freeze-in yields correspond to smaller $\xi$, we find that increasing the freeze-in yield also increases the final DM abundance after dark sector freeze-out. 

\providecommand{\href}[2]{#2}\begingroup\raggedright\endgroup

\end{document}